\documentclass[12pt,a4paper]{lmcs}

\usepackage{dsfont}
\usepackage{nicematrix}
\usepackage{comment}
\usepackage[normalem]{ulem} 
\usepackage{xcolor}

\newcommand{\grammar}{\ensuremath{\Lambda}}
\newcommand{\dee}{\mathrm{d}}
\newcommand{\cA}{\mathcal{A}}
\newcommand{\cC}{\mathcal{C}}
\newcommand{\cE}{\mathcal{E}}
\newcommand{\cF}{\mathcal{F}}
\newcommand{\cG}{\mathcal{G}}
\newcommand{\cH}{\mathcal{H}}
\newcommand{\cL}{\mathcal{L}}
\newcommand{\cM}{\mathcal{M}}
\newcommand{\cO}{\mathcal{O}}
\newcommand{\cP}{\mathcal{P}}
\newcommand{\cX}{\mathcal{X}}
\begin{document}

\title{Two behavioural pseudometrics for continuous-time Markov processes}

\author[L.~Chen]{Linan Chen}[a]
\author[F.~Clerc]{Florence Clerc}[b]
\author[P.~Panangaden]{Prakash Panangaden}[a]

% affiliation 1 (automatically numbered a)
\address{McGill University}	%optional

% affiliation 2 (automatically numbered b)
\address{Heriot-Watt University}	%optional

%\author{Linan Chen (McGill University) \footnote{Linan Chen was supported by an NSERC discovery grant} \and
%Florence Clerc (Heriot-Watt University)\footnote{Florence Clerc was supported by the EPSRC-UKRI grant EP/Y000455/1, by a CREATE grant for the project INTER-MATH-AI, by NSERC and by IVADO through the DEEL Project CRDPJ 537462-18} \and
%Prakash Panangaden (McGill University)\footnote{Prakash Panangaden was supported by an NSERC discovery grant}}

\begin{abstract}
Bisimulation is a concept that captures behavioural equivalence of states in a variety of types of transition systems.  It has been widely studied in discrete-time settings where a key notion is the bisimulation metric which quantifies ``how similar two states are''. 

In our previous FoSSaCS paper,  we generalized the concept of bisimulation metric in order to metrize the behaviour of continuous-time Markov processes. Similarly to the discrete-time case, we constructed a pseudometric following two iterative approaches -- through a functional and through a real-valued logic, and showed that the outcomes coincide: 
the pseudometric obtained from the logic is a specific fixpoint of the functional which yields our first pseudometric. %the obtained pseudometric being a fixpoint in both approaches.
However, unlike the discrete-time setting, in which the process has a step-by-step dynamics, the behavioural pseudometric we constructed applies to Markov processes that evolve continuously through time, such as diffusions and jump diffusions.  
%The whole discrete-time approach relies entirely on the step-based dynamics: the process jumps from state to state. We define a behavioural pseudometric for processes that evolve continuously through time, such as Brownian motion or involve jumps or both.

While our treatment of the pseudometric in the FoSSaCS paper relied on the time-indexed Markov kernels, our earlier works on behavioural equivalences highlighted the importance of trajectories for true continuous-time Markov processes. Building on these developments, in this paper we propose a second
behavioural pseudometric for diffusions based on trajectories. We conduct a similar study of this pseudometric from both the perspective of a functional and the viewpoint of a real-valued logic. We also compare this pseudometric with the first pseudometric obtained in the FoSSaCS paper.
\end{abstract}

\maketitle

\section{Introduction}
Bisimulation~\cite{Milner80,Park81,Sangiorgi09} is a fundamental concept in the theory of transition systems capturing a strong notion of behavioural equivalence.  The extension to probabilistic systems is due to Larsen and Skou~\cite{Larsen91}; henceforth we will simply say ``bisimulation'' instead of ``probabilistic bisimulation''.  Bisimulation has been studied for discrete-time systems where transitions happen as steps, both on discrete~\cite{Larsen91} and continuous state spaces~\cite{Blute97,Desharnais98,Desharnais02}.  In all these types of systems, a crucial ingredient of the definition of bisimulation is the ability to talk about \emph{the next step}.  This notion of bisimulation is characterized by a modal logic~\cite{Larsen91} even when the state space is continuous~\cite{Desharnais98}.

Previous work on discrete-time Markov processes by Desharnais et al.~\cite{Desharnais99b,Desharnais04}
extended the modal logic characterizing bisimulation to a real-valued logic that allowed
to not only state if two states were ``behaviourally equivalent'' but, more interestingly, how similarly they behaved.  This shifts the notion from a qualitative notion (an
equivalence) to a quantitative one (a pseudometric).

Other work also on discrete-time Markov processes by van Breugel et
al.~\cite{vanBreugel05a} introduced a slightly different real-valued logic and compared
the corresponding pseudometric to another pseudometric obtained as a terminal coalgebra of a
carefully crafted functor. We also mention in this connexion the work by Ferns et al. on Markov Decision Processes and the connexion between bisimulation and optimal value functions \cite{Ferns11}.

\vspace{0.5cm}

While bisimulation for discrete-time systems has been extensively studied, the counterpart literature in the continuous-time setting is still scarce. Since a growing part of what computer science allows us to do takes place in real-time (e.g., robotics, self-driving cars, and online machine-learning), it is becoming increasingly important to develop a way to quantify behaviours of systems that evolve continuously in time.

Some previous work has been done on the so called continuous-time Markov chain (see for example~\cite{Baier08}), but such a process takes values in a finite state space and hence still possesses a discrete-time \emph{step} in nature; it is only the real-valued time duration associated with each state that leads to the name ``continuous-time''. Such processes are known as ``jump processes'' in the mathematical literature (see, for example, \cite{Rogers00a,Whitt02}), a phrase that better captures the true nature of such processes.  Metrics and equivalences for such processes were studied by Gupta et al.~\cite{Gupta04,Gupta06} and more recently by Bacci et al.~ \cite{Bacci17}.

The processes we consider have continuous state spaces and are governed by continuous-time evolutions, where a paradigmatic example is the Brownian motion. A common approach to tackle these continuous-time processes is to discretize time, but it is well-known that this procedure generates errors that may accumulate over time and lead to vastly different outcomes. In addition, when approximating continuous-time processes by discrete-time ones, entirely new phenomena and difficulties arise.  For example, even the basic properties of trajectories of the Brownian motion are significantly more complicated than the counterparts of a random walk.  Basic concepts like ``the time at which a process exits a given subset of the state space'' become intricate to define.  Notions like ``matching transition steps'' are no longer applicable as the notion of ``step'' does not make sense. 

\subsection*{Overview of our work} In~\cite{Chen19a,Chen20,Chen23}, we proposed different notions of behavioural
equivalences for continuous-time processes. We showed that there were several possible extensions of the notion of bisimulation to the continuous-time setting, and that the continuous-time notions needed to involve trajectories in order to be meaningful.  There were significant mathematical challenges in even proving that an equivalence relation existed.  For example,  obstacles occurred in establishing measurability of various functions and sets, due to the inability to countably generate the relevant $\sigma$-algebras. 
%Those papers left completely open the question of defining a suitable pseudometric analogue, a concept that would be more useful in practice than an equivalence relation.

In~ \cite{Chen25a} and this work, we take a step further in extending the behavioural study from discrete-time settings to continuous-time cases. We tackle the quantitative analogue of behavioural equivalences by constructing two behavioural pseudometrics for continuous-time Markov processes: one is based on a time-indexed family of  Markov kernels (the work conducted in \cite{Chen25a}), and the other is derived from a space-indexed family of probability measures on the space of trajectories (the work conducted in this paper). Moreover, each of these pseudometrics is defined in two different ways.
\begin{enumerate}
\item The first way of defining a pseudometric is through a functional. We
define a functional on a certain class of pseudometrics that ``deforms'' pseudometrics according to the dynamics of the process.  More
specifically, viewing a pseudometric as a cost function, the said functional is just
the optimal transport cost between probability measures associated with the process. Then, we obtain the target behavioural pseudometric through iteratively applying the functional; we also confirm that this pseudometric is indeed a fixpoint of the functional.
\item We then show that the pseudometric acquired above is characterized by a real-valued logic that closely resembles the one introduced in
\cite{vanBreugel05a} for discrete-time systems:  this is a quantitative analogue of the logical characterization of
bisimulation.
\end{enumerate}
We further establish a comparison result between the two behavioural pseudometrics and use a concrete example to illustrate our approaches and results.  

Very broadly speaking, we follow a familiar path from equivalences to logics to metrics.  However, it is necessary for us to redevelop the framework and the mathematical techniques from scratch. 
Indeed, a very important aspect in discrete-time is the fact that the process is a jump process, ``hopping'' from state to state. In the continuous-time setting, there are new mathematical challenges that need to be overcome.
This means that the similarity between the pre-existing work on discrete-time systems and our generalization to continuous-time models is only at the highest level of abstraction.

%In~ \cite{Chen25a}, the pseudometric that we defined relied on the dynamics of the system being described by a time-indexed family of Markov kernels. However, we have shown in~ \cite{Chen19a,Chen20,Chen23} that with very loose conditions on the processes we considered, trajectories had to be involved in order to define meaningful notions of behavioural equivalences. In this work, we will adapt our work from ~\cite{Chen25a} to trajectories. An example will enable us to compare the two approaches.

\subsection*{Outline of the paper}
In Section \ref{sec:background}, we first review some measure theoretical notions and then introduce the model of continuous-time Markov process under investigation in this work. Since this work adapts our previous work from \cite{Chen25a}, in Section \ref{sec:previous-work} we briefly recall the construction of the first pseudometric $\overline{\delta}$ and the important results on $\overline{\delta}$. In Section \ref{sec:functional}, we introduce a functional $\cG$ and define the second pseudometric $\overline{d}$ via $\cG$; we also show that $\overline{d}$ is a fixpoint of $\cG$. In Section \ref{sec:logicG}, we prove that this pseudometric $\overline{d}$ is also characterized by a real-valued logic. We will then draw a comparison to the work in \cite{Chen25a} and establish an order relation between the two pseudometrics $\overline{\delta}$ and $\overline{d}$ in Section \ref{sec:partial-conclusion}.
In Section \ref{sec:example}, we further re-examine one of our examples from \cite{Chen25a} under both pseudometrics. 
Finally, in Section~\ref{sec:conclusion} we discuss the limitations of our approach and how it relates to previous works.

\section{Mathematical background}
\label{sec:background}

\subsection{Background on measure theory} We assume the reader to be familiar with basic measure
theory and topology. Nevertheless we provide a brief review of the relevant notions and theorems. Let us start with clarifying a few notations on integrals: Given a measure space $(X,\Sigma_X,\mu)$ and a measurable function $f: X \to \mathbb{R}$, 
we write either $\int f(x) ~ \mu (\dee x)$ or $\int f~ \dee \mu$ interchangeably for the integral of $f$ under $\mu$ (whenever it exists). The second notation will be especially useful when considering a Markov kernel $P_t(x,\cdot)$ for some $t \geq 0$ and $x \in X$: $\int f(y) ~P_t(x, \dee y) = \int f ~ \dee P_t(x)$.

   \subsubsection{Lower semi-continuity}

\begin{defi}
Given a topological space $X$, a function $f: X\to \mathbb{R}$ is \emph{lower semi-continuous} if for every $x_0 \in X$, $\liminf_{x \rightarrow x_0} f(x) \geq f(x_0)$. This condition is equivalent to the following: for any $y \in \mathbb{R}$, the set $f^{-1}((y, + \infty)) := \{ x ~|~ f(x) > y\}$ is open in $X$.
\end{defi}

Let $X$ be a metric space. A function $f: X \to \mathbb{R}$ is lower semi-continuous if and only if $f$ is the limit of an increasing sequence of real-valued continuous functions on $X$, by Baire's approximation theorem (see
\cite[p.~71]{Mustata08}).

    \subsubsection{Couplings}
\label{sec:coupling}

\begin{defi}
Let $(X, \Sigma_X, P)$ and $(Y, \Sigma_Y, Q)$ be two probability spaces.  Then a \emph{coupling} $\gamma$ of $P$ and $Q$ is a probability measure  on $(X \times Y, \Sigma_X \otimes \Sigma_Y)$ such that for every  $B_X \in \Sigma_X$, $\gamma(B_X \times Y) = P(B_X)$ and for every $B_Y \in \Sigma_Y$, $\gamma(X \times B_Y) = Q(B_Y)$ ($P, Q$ are called the \emph{marginals} of $\gamma$).  We write $\Gamma(P, Q)$ for the set of couplings of $P$ and~$Q$.
\end{defi}

In other words, a coupling is a joint distribution on the space $X\times Y$ while preserving the prescribed marginal distributions $P$ and $Q$ in the respective coordinates. In this paper, $X$ and $Y$ will be either the
state space (at a given time) or the space of trajectories. A pseudometric
assigns a cost to each pair in $X\times Y$, and minimizing its expectation over all couplings yields the optimal transport cost
used to compare the two distributions $P$ and $Q$.

\begin{restatable}{lem}{lemmacouplingscompact}
\label{lemma:couplings-compact}
Given two probability measures $P$ and $Q$ on Polish spaces $X$ and $Y$ respectively,  the set of
couplings $\Gamma (P, Q)$ is compact under the topology of weak convergence. 
\end{restatable}

\textcolor{black}{To verify the above assertion, we first note that the family of couplings with fixed marginals is tight because of the following simple argument: for compact sets $K\subseteq X$ and $L\subseteq Y$,
\[
 \gamma((K\times L)^c)\leq P(K^c)+Q(L^c)
 \quad\text{for every }\gamma\in\Gamma(P,Q).
\]
Then, it follows from Prokhorov's theorem that $\Gamma(P,Q)$ is relatively compact, with the weak limits retaining the prescribed marginals, and hence the family is closed and compact.}

   \subsubsection{Optimal transport theory}

An important part of this work relies on the optimal transport theory. We adapt relevant contents in \cite{Villani08} to our framework.

\begin{comment}
Consider a Polish space $\cX$ and a lower semi-continuous \emph{cost function} $c: \cX \times \cX \to [0,1]$ such that for every $x \in \cX$, $c(x,x) = 0$. For every two probability measures $\mu$ and $\nu$ on $\cX$, we write $W(c)(\mu, \nu)$ for the \emph{optimal transport cost} from $\mu$ to $\nu$.
Adapting Theorem 5.10(iii) of \cite{Villani08} to our framework,  we get the following statement for the \emph{Kantorovich duality}:
\[W(c)(\mu, \nu) :=  \min_{\gamma \in \Gamma(\mu, \nu)} \int c ~ \dee \gamma
= \max_{h \in \cH(c)} \left| \int h ~ \dee \mu - \int h ~\dee \nu  \right| \]
where $ \cH(c) := \{ h: \cX \to [0,1] ~|~ \forall x,y\in\cX, ~~ |h(x) - h(y)| \leq c(x,y) \}$.
   
\begin{restatable}{lem}{lemmaoptimaltransportpseudometric}
%\label{lemma:optimal-transport-pseudometric}
If the cost function $c$ is a 1-bounded  pseudometric on $\cX$, then $W(c)$ is a 1-bounded pseudometric on the space of probability measures on $\cX$.
\end{restatable}
\end{comment}

Consider a Polish space $\cX$ and a lower semi-continuous $1$-bounded\footnote{Here, $1$-bounded means that $c(x,y)\leq1$ for all $x,y\in\mathcal X$.} pseudometric $c: \cX \times \cX \to [0,1]$. As we will discuss in Section \ref{sec:functional}, this pseudometric $c$ does not need to generate
the background topology, but the lower semi-continuity guarantees that $c$ is Borel measurable for the background topology.

For every two probability measures $\mu$ and $\nu$ on $\cX$, we write $W(c)(\mu, \nu)$ for the \emph{optimal transport cost} from $\mu$ to $\nu$. The pseudometric is called the \emph{cost function}.
Adapting Theorem 5.10(iii), Particular Case 5.4 and Particular Case 5.16 of \cite{Villani08} to our framework, we have the following statement for the \emph{Kantorovich duality}:
\[W(c)(\mu, \nu) :=  \min_{\gamma \in \Gamma(\mu, \nu)} \int c ~ \dee \gamma
= \max_{h \in \cH(c)} \left| \int h ~ \dee \mu - \int h ~\dee \nu  \right| \]
where $ \cH(c) := \{ h: \cX \to [0,1]\text{ Borel measurable} ~|~ \forall x,y\in\cX, ~~ |h(x) - h(y)| \leq c(x,y) \}$.

Moreover, $W(c)$ is again a $1$-bounded pseudometric on the space of probability measures on $\cX$. Indeed, $W(c)$ is clearly $1$-bounded and symmetric, and given three probability measures $\mu, \nu,\theta$ on $\cX$, for any $h\in\cH(c)$, \[
\left|\int h\dee\mu-\int h\dee\theta\right|\leq\left|\int h\dee\mu-\int h\dee\nu\right|+\left|\int h\dee\nu-\int h\dee\theta\right|\leq W(c)(\mu,\nu)+W(c)(\nu,\theta),
\]
which implies that $W(c)$ satisfies the triangle inequality. \\

\color{black}

We will later need the following technical lemma. Theorem 5.20 of \cite{Villani08} states that a sequence $W(c_k)(P_k, Q_k)$ converges to $W(c)(P, Q)$ if $c_k$ uniformly converges to $c$ and $P_k$ and $Q_k$ converge weakly to $P$ and $Q$ respectively. Uniform convergence in the cost function may be too strong a condition for us, but the following lemma is enough for what we need.

\begin{restatable}{lem}{lemmawassersteinlimit}
\label{lemma:wasserstein-limit}
Consider a Polish space $\cX$ and a 1-bounded pseudometric $c: \cX \times \cX \to [0,1]$ such that there exists an increasing ($c_{k+1} \geq c_k$ for every $k$) sequence of continuous \textcolor{black}{(when viewed as functions)} pseudometrics $c_k: \cX \times \cX \to [0,1]$ that converges to $c$ pointwise.  Then, given two probability measures $P$ and $Q$ on $\cX$,
\[ \lim_{k \rightarrow \infty} W(c_k)(P,Q) = W(c)(P, Q). \]
\end{restatable}

\begin{proof}
For each $c_k$, the optimal transport cost $W(c_k)(P, Q)$ is attained for a coupling $\pi_k$.  Using Lemma \ref{lemma:couplings-compact}, we know that the space of couplings $\Gamma(P, Q)$ is compact. We can thus extract a subsequence that we will still denote by $(\pi_k)_{k \in \mathbb{N}}$ which converges weakly to some coupling $\pi \in \Gamma(P, Q)$. We will show that this coupling $\pi$ is in fact an optimal transference plan, i.e. it achieves the minimum value: $W(c)(P, Q) :=  \min_{\gamma \in \Gamma(P, Q)} \int c ~ \dee \gamma = \int c ~ \dee \pi$.

By Monotone Convergence Theorem,
\[ \int c ~ \dee \pi = \lim_{k \rightarrow \infty } \int c_k ~ \dee \pi. \]
Consider $\epsilon > 0$. There exists $k$ such that
\begin{align}
\label{eq:numberg}
\int c ~ \dee \pi & \leq \epsilon + \int c_k ~ \dee \pi
\end{align}

Since $\pi_k$ converges weakly to $\pi$ and since $c_k$ is continuous and bounded,
\[ \int c_k ~ \dee \pi = \lim_{n \rightarrow \infty} \int c_k ~ \dee \pi_n,\]
which implies that there exists $n_k \geq k$ such that
\begin{align}
\label{eq:numberh}
\int c_k ~ \dee \pi & \leq \epsilon +  \int c_k ~ \dee \pi_{n_k}.
\end{align}
Putting Equations (\ref{eq:numberg}) and (\ref{eq:numberh}) together, we get
\begin{align*}
\int c ~ \dee \pi
& \leq 2 \epsilon +  \int c_k ~ \dee \pi_{n_k}\\
&  \leq 2 \epsilon +  \int c_{n_k} ~ \dee \pi_{n_k} \quad \text{ since $(c_k)_{k \in \mathbb{N}}$ is increasing}\\
 & = 2 \epsilon + W(c_{n_k})(P, Q).
\end{align*}
This implies that
\[ \int c ~ \dee \pi \leq \lim_{k \rightarrow \infty} W(c_k)(P, Q). \]

The other inequality is trivial since $c_k \leq c$.
\end{proof}

\subsection{Background on Feller-Dynkin process}
\label{sec:CTsystems}

This work focuses on continuous-time processes that are honest (without loss of mass over time) and with additional regularity conditions.  In order to define what we mean by continuous-time Markov processes here, we first define Feller-Dynkin processes.  Much of this material is adapted from \cite{Rogers00a} and we use
their notations.  Another useful source is \cite{Bobrowski05}.  

\subsubsection{Feller-Dynkin semigroup}
Let $E$ be a locally compact, Hausdorff
space with a countable base.  We also equip the set $E$ with its Borel $\sigma$-algebra $\mathcal{B}(E)$, denoted by $\cE$.  The previous topological hypotheses also imply that $E$ is $\sigma$-compact and Polish (see corollary IX.57 in \cite{Bourbaki89b}).
We denote by $\Delta$ a 1-bounded metric that generates the topology making $E$ Polish.

\begin{defi}
A \emph{semigroup} of operators on any Banach space $X$ is a family
of linear continuous (bounded) operators $\cP_t: X \to X$ indexed by
$t\in\mathbb{R}_{\geq 0}$ such that
\[ \forall s,t \geq 0, \cP_s \circ \cP_t = \cP_{s+t} \qquad \text{(semigroup property)}\]
and
\[ \cP_0 = I \qquad \text{(the identity)}.  \]
\end{defi}
\begin{defi}
For $X$ a Banach space, we say that a semigroup $\cP_t:X\to X$ is \emph{strongly continuous} if 
\[ \forall x\in X,\qquad \lim_{t\downarrow 0}\| \cP_t x - x \|_X = 0.  \]
\end{defi}

What the semigroup property expresses is that we do not need to understand the past (what
happens before time $t$) in order to compute the future (what happens after some
additional time $s$, so at time $t+s$) as long as we know the present (at time
$t$). 

We say that a continuous real-valued function $f$ on $E$ \emph{vanishes at
infinity} if for every $\varepsilon > 0$ there is a compact subset
$K \subseteq E$ such that for every $x\in E\setminus K$, we have
$|f(x)| \leq \varepsilon$.  To give an intuition, if $E$ is the real line, this means that
$\lim_{x \rightarrow \pm \infty} f(x) = 0$.  The space $C_0(E)$ of continuous real-valued functions
that vanish at infinity is a Banach space with the ``$\sup$'' norm.   
\begin{defi}
A \emph{Feller-Dynkin (FD) semigroup} is a strongly continuous
semigroup $(\hat{P}_t)_{t \geq 0}$ of linear operators on $C_0(E)$ satisfying the
additional condition:  
\[\forall t \geq 0 ~~~ \forall f \in C_0(E) \text{, if }~~ 0 \leq f \leq 1 \text{, then }~~ 0 \leq \hat{P}_t f \leq 1.\]
\end{defi}

The Riesz representation theorem can be found as Theorem II.80.3 of \cite{Rogers00a}.  From it, we can derive the following important proposition which relates these FD-semigroups with Markov
kernels (see \cite{Williams91}).  This allows one to see the connection with familiar
probabilistic transition systems.  

\begin{prop}
\label{prop:Riesz-use}
Given an FD-semigroup $(\hat{P}_t)_{t \geq 0}$ on $C_0(E)$, it is possible to define a
unique family of (sub-)Markov 
kernels $(P_t)_{t \geq 0} : E \times \mathcal{E} \to [0,1]$ such that for
all $f \in C_0(E)$, 
\[\forall t\geq 0\;\forall x\in E,\quad \hat{P}_t f(x) = \int f(y) P_t(x, \dee y).  \]
\end{prop}
Given a time $t$ and a state $x$, we will often write $P_t(x)$ for the (sub-)probability measure $P_t(x, \cdot)$ on $E$. Note that since $E$ is Polish, $P_t(x)$ is tight. 

In this work, we focus on FD-semigroups that are \emph{honest}; intuitively speaking, an honest FD-semigroup has no loss of mass over time.

\begin{defi}
\label{def:honest}
An FD-semigroup $(\hat{P}_t)_{t \geq 0}$ is \emph{honest} if for every $x \in E$ and every time $t \geq 0$, $P_t(x, E) = 1$; that is, $(P_t)_{t\geq 0}$ are Markov kernels on $E\times\cE$.
\end{defi}

From now on, we will always assume the concerned FD-semigroup to be honest. However, we want to point out that this work can be adapted to non-honest semigroups with some additional continuity condition on what an outside agent observes (this will be introduced later).

\subsubsection{Trajectories}
\label{sec:trajectories}

Compared to our previous work in \cite{Chen25a}, trajectories are the new ingredient that we are introducing here based on the intuition discussed in \cite{Chen19a,Chen20,Chen23}. The space of trajectories does not
appear explicitly in the study of the labelled Markov processes but one does
see it in the study of continuous-time Markov chains and jump processes.

We will impose further conditions (see Definition \ref{def:diffusion}), but for now we follow \cite{Rogers00a} to introduce the concept of trajectories and the corresponding probabilities.

\begin{defi}
A function $\omega : [0,\infty)\to E$ is called a \emph{trajectory on $E$} if it is \emph{c\`adl\`ag} \footnote{c\`adl\`ag
  stands for the French ``continu \`a droite, limite \`a gauche''} meaning that for every $t \geq
0$, $\lim_{s > t, s \to t} \omega(s) = \omega (t)$ and for every $t > 0$, $ \omega(t-) := \lim_{s < t, s \to t} \omega(s)$ exists.
\end{defi}
As an intuition, a c\`adl\`ag function $\omega$ is an ``almost continuous'' function with
jumping in a reasonable fashion.  

It is possible to associate to such an FD-semigroup a \emph{canonical FD-process}.
Let $\Omega$ be the set of all trajectories $\omega : [0, \infty ) \to
E$. 
\begin{defi}
\label{def:canonical-FD-process}
The
\emph{(canonical) FD-process} associated to an FD-semigroup $(\hat{P}_t)_{t \geq 0}$ is
\[(\Omega, \cA, (\cA_t)_{t \geq 0}, (X_t)_{t \geq 0}, (\mathbb{P}^x)_{x \in E})\]
where
\begin{itemize}
\item the random variable $X_t: \Omega \to E$ is defined as $X_t(\omega) := \omega (t)$ for every $\omega\in\Omega$ and $t\geq 0$,
\item $\cA := \sigma (X_s ~|~ s\geq 0)$ \footnote{The $\sigma$-algebra $\mathcal{A}$ is the same as the one induced by the Skorohod metric, see theorem 16.6 of \cite{Billingsley99}}, $\cA_t := \sigma (X_s ~|~ 0 \leq s \leq t)$ for every $t\geq 0 $,
\item  given any probability measure $\mu$ on $E$,
by the existence theorem for canonical Feller--Dynkin
processes \cite[Chapter~III, Section~7]{Rogers00a},
there exists a unique probability measure
$\mathbb{P}^{\mu}$ on $(\Omega,\mathcal{A})$ such that for all
  $n \in \mathbb{N}, 0 < t_1 < t_2 < ...  < t_n$ and
  $x_0, x_1, ..., x_n$ in $E$,
\begin{align*}
   \mathbb{P}^\mu (X_0 \in dx_0, X_{t_1} \in dx_1, ..., X_{t_n} \in dx_n) 
   =
  \mu (dx_0) P_{t_1}(x_0, dx_1)...P_{t_n -
    t_{n-1}}(x_{n-1}, dx_n)
  \end{align*}
 The $dx_i$ in this equation should be understood as infinitesimal volumes.  This notation is standard in probability and should be understood by integrating it over measurable state sets $C_0,C_1,...,C_n$. We write $\mathbb{P}^x = \mathbb{P}^{\delta_x}$, where $\delta_x$ is the Dirac delta distribution at $x\in E$. Clearly, the
measure $\mathbb{P}^x$ is the distribution of the FD process starting at the point~$x$.
\end{itemize}
\end{defi}

We will also call an FD-process \emph{honest} if it is associated to an honest FD-semigroup.
\begin{rem}
Here we have introduced an FD-process to which we refer as the canonical process.  Note that there could be other tuples
\[(\Omega', \mathcal{A}', (\cA'_t)_{t \geq 0}, (Y_t)_{ t \geq 0}, (\mathbb{Q}^x)_{x \in E})\]
satisfying the same conditions except that the sample space $\Omega'$ may no longer be the trajectory space.
\end{rem}

\begin{rem}
Let us give an intuition for this new way of describing dynamics in the context of discrete-time Markov chains.  One standard way of describing such a system is through the Markov kernel $\tau$, where the analogue in the continuous-time setting is the time-indexed family of Markov kernels $(P_t)_{t \geq 0}$ on the state space. On the other hand, it is also possible to look at infinite runs of the process. For instance, for a coin toss, we can look at sequences of heads and tails.  Such individual sequences correspond to trajectories and we can look at the probability of certain sets of trajectories (for instance, the set of trajectories that start with three heads). In the continuous-time setting, these probabilities are governed by the space-indexed family of measures $(\mathbb{P}^x)_{x \in E}$ on the trajectory space. 
\end{rem}

\subsubsection{Diffusions}
\label{sec:diffusions}
This part is the preparation for the work in Section \ref{sec:functional} and \ref{sec:logicG}. 

In the case when the trajectories of an FD-process are continuous in time, we may assume that $\Omega$ is the space of continuous functions
$\mathbb{R}_{\geq0}\rightarrow E$, equipped with the Skorohod topology $J_1$, and then $\Omega$ is a Polish space, because the space of continuous maps from a locally-compact Polish space into a Polish space is Polish. Since $\Omega$ consists of continuous functions, the Skorohod topology $J_1$ is also equivalent to the compact-open topology.

\begin{defi}
\label{def:diffusion}
A \emph{diffusion} is an honest FD-process satisfying the following properties:
\begin{itemize}
\item the trajectories are continuous in time and hence, as stated above, we will take $$\Omega:=\{\omega:\mathbb{R}_{\geq0}\rightarrow E\textup{ is continuous}\}.$$
%\item for every state $x$, the probability measure $\mathbb{P}^x$ is tight (it is also called ``inner regular''): for every $\epsilon > 0$, there exists a compact $K$ such that $\mathbb{P}^x (K) > 1 - \epsilon$, and
\item the map $x \mapsto \mathbb{P}^x$ is weakly continuous meaning that if $x_n \underset{n \rightarrow \infty}{\longrightarrow} x$ in $E$, then $\mathbb{P}^{x_n}\underset{n \rightarrow \infty}{\Longrightarrow}\mathbb{P}^x$, where ``$\Longrightarrow$'' refers to the weak convergence of probability measures in the sense that, for every bounded and continuous function $f$ on $\Omega$, $\int f \dee \mathbb{P}^{x_n} \underset{n \rightarrow \infty}{\longrightarrow} \int f \dee \mathbb{P}^{x}$.
\end{itemize}
\end{defi}

In \cite{Rogers00a},  FD-diffusions are also introduced. Our definition is different in the sense that we only define them for honest processes, but we require fewer conditions. This model of diffusion covers a wide range of stochastic processes, with the most classical example being the Brownian motion that we will introduce below. 
\medskip

\noindent\textbf{Brownian motion as an example:} \textcolor{black}{Take $E$ to be the real line $\mathbb R$ with its usual topology. Given $x\in\mathbb R$, the one-dimensional \emph{Brownian motion starting from $x$} is  $(X_t^x:=x+B_t)_{t\geq0}$ (where $(B_t)_{t\geq0}$ is the standard Brownian motion starting from $0$); this is an honest FD-process with continuous trajectories, and its Markov kernel is \[
P_t(x,\dee y):=\frac 1{\sqrt{2\pi t}}\exp\left(-\frac {|x-y|^2}{2t}\right)\dee y\;\text{ for }t>0, \text{ with }P_0(x)=\delta_x.
\]
We will use this Brownian motion model to illustrate some of our results in later sections. For a comprehensive exposition on the Brownian motion, we refer readers to \cite{Karatzas12}.}
\medskip

The work conducted in Section \ref{sec:functional} and \ref{sec:logicG}, which is the construction and the study of the second behavioural pseudometric, is restricted to diffusions. Note that the weak continuity condition that we impose on $x\mapsto\mathbb{P}^x$ for diffusions implies that the map $x\mapsto P_t(x)$ is also weakly continuous for every $t$; the latter is the condition adopted for the process treated in \cite{Chen25a}, which we will review in the next section.

  \subsection{Observables}

In previous sections,  we defined Feller-Dynkin processes.  In order to bring the processes more in line
with the kind of transition systems that have hitherto been studied in the computer
science literature, we also equip the state space $E$ with an additional continuous function
$obs: E \to [0,1]$, to which we refer as the \emph{observable} function. 
One should think of it as the interface between the process and the
user (or an external observer): external observers won't see the exact state in which the process
is at a given time, but they will see the associated observables. For example, a diver's depth gauge reveals the depth
without revealing the diver's exact underwater location.

\textcolor{black}{Here is a concrete example of an observable function: taking again $E=\mathbb R$ with the usual topology, define $obs(x):=\max\{0,1-|x|\}$ for every $x\in\mathbb R$. Clearly, $obs:\mathbb R\to[0,1]$ is $1$-bounded and continuous; it models a sensor for which the signal is the strongest at the origin and decays linearly to zero at distance $1$ (without telling from which side of the origin the signal originates), beyond which it provides no information about positions.}

Note that the continuity condition on the observable is the same as in \cite{Chen25a} but is a major difference from our previous work \cite{Chen19a,Chen20} where we used a countable set of atomic propositions $AP$ and $obs$ was a discrete function $E \to 2^{AP}$.

\section{A pseudometric based on Markov kernels}
\label{sec:previous-work}

The approach that we are now following is derived from our previous work in \cite{Chen25a} but adapted to trajectories. Let us first summarize this approach so that we are then ready to tackle the added difficulties posed by trajectories. All the details for this section can be found in \cite{Chen25a}.

In \cite{Chen25a}, we adopt the following model of FD-process to construct the first behavioural pseudometric. 
\begin{defi}
    A \emph{Continuous-Time Markov process} (abbreviated CTMP) is the FD-process associated with an honest FD-semigroup on $C_0(E)$, equipped with a continuous observable function $obs: E \to [0,1]$, and satisfies that if  $x_n\underset{n\rightarrow\infty}{\longrightarrow}x$ in $E$, then $
    P_t(x_n)\underset{n\rightarrow\infty}{\Longrightarrow} P_t(x)
    $ for every $t$,
    where ``$\Longrightarrow$'' again refers to the weak convergence of probability measures.
\end{defi}
\subsection{Through a functional}

At the core of our construction is the definition of a functional on the sets of 1-bounded pseudometrics on the state space that will admit a fixpoint. 

We need to consider three classes of pseudometrics in order to define our functional:
\begin{enumerate}
    \item $\cM$ is the class of 1-bounded pseudometrics on the state space $E$ equipped with the order ``$\leq$'' defined as: $m_1 \leq m_2$ if and only if for every $(x,y)$, $m_1(x,y) \leq m_2(x,y)$.
    \item We then define a subset $\cP$ of $\cM$ by restricting to pseudometrics that are lower semi-continuous (with respect to the original topology $\cO$ on $E$ generated by the metric $\Delta$ that makes $E$ Polish).
    \item Finally, let $\cC$ be the class consisting of pseudometrics $m \in \cM$ such that the topology  generated by $m$ on $E$ is a subtopology of the original topology $\cO$, \emph{i.e.}\  $m$ is a continuous function $E \times E \to [0,1]$.
\end{enumerate}

\noindent \textcolor{black}{\textbf{An important distinction on pseudometrics and the topology:}} It is very important to note that the topology $\cO$ on $E$ is generated by the 1-bounded metric $\Delta$, and hence $\Delta$ is in $\cC$. However, we can define many pseudometrics that are not necessarily continuous with respect to $\cO$. \textcolor{black}{ Let us illustrate this point with a simple example. Assume $E$ is the real line $\mathbb{R}$ and $\mathcal{O}$ is the usual topology generated by the $1$-bounded version of the Euclidean metric $\Delta(x,y):=\min\{1,|x-y|\}$ for $x,y\in\mathbb R$. Define the \emph{discrete metric} on $\mathbb R$ to be $m(x, y) :=\mathds{1}_{x\neq y} $ for $x,y\in\mathbb{R}$. Then, $m$ is lower semi-continuous with respect to $\mathcal O$, since for any $r\in\mathbb R$, the set $m^{-1}((r,\infty)):=\{(x,y)\in\mathbb R^2:m(x,y)>r\}$ is either $\mathbb R^2$, or $\mathbb R^2\setminus\{(x,x):x\in\mathbb R\}$, or $\emptyset$, all of which are open, and hence $m$ belongs to $\cP$. However, $m\notin\cC$, as $m(0,\frac 1n)=1$ for all $n\geq1$, while $\Delta(0,\frac1n)\to0$ as $n\to\infty$, which means $m$ is not continuous.  
%The usual topology on $\mathbb{R}^2$ is generated by the bounded Euclidean metric defined as $d_{\text{E}} ((x_1, x_2), (y_1, y_2)) = \min \left\{ 1, \sqrt{(x_1 - y_1)^2 + (x_2 - y_2)^2} \right\}$. This topology is also generated by the bounded Manhattan distance defined as $d_{\text{M}} ((x_1, x_2), (y_1, y_2)) = \min \left\{ 1, |x_1 - y_1| + |x_2 - y_2|\right\}$. The discrete metric however does not relate to the standard topology on $\mathbb{R}^2$.
} 
In general, we have the following inclusion: $\cC \subseteq \cP \subseteq \cM$.

The pointwise supremum of any family of $1$-bounded pseudometrics is again a $1$-bounded pseudometric. If each member of the family is lower semi-continuous, then so is its supremum. In particular, the pointwise supremum of an increasing sequence in $\mathcal C$ belongs to $\mathcal P$, although it need not belong to $\mathcal C$. 

Consider $(P_t)_{t\geq0}$ a family of Markov kernels associated with a CTMP.
Given a \emph{discount factor} $0 < c\le1$,  we define the functional $\cF_c : \cP \to \cM$ as follows: for every pseudometric $m \in \cP$ and every two states $x,y\in E$,
\[ \cF_c(m)(x,y) := \sup_{t \geq 0} c^t W(m) (P_t(x), P_t(y)).\]
The pseudometric $ \cF_c(m)$ applied to the states $x$ and $y$ compares the probability measures $P_t(x)$ and $P_t(y)$  through transport theory and takes their supremum. We allow $c=1$ in this definition, corresponding to
the undiscounted case. The subsequent continuity
and iterative fixed-point results are stated for
$0<c<1$.

Now fix $0<c<1$. While we do not know if $\cF_c(m)$ is lower semi-continuous for general $m \in \cP$, we are able to show that if $m \in \cC$, then $\cF_c(m) \in \cC$. This enables us to define an increasing sequence of pseudometrics in $\cC$: for every $x,y\in E$,
\begin{align*}
\delta^c_0(x,y) & := |obs(x) - obs(y)|,\\
\delta^c_{n+1}(x,y) & := \cF_c(\delta_n^c)(x,y)\;\textup{ for }n\geq 0.
\end{align*}
Then we define the pseudometric $\overline{\delta}^c := \sup_{n\geq1} \delta^c_n$ (which is also a limit since the sequence is non-decreasing). As the supremum of a sequence of continuous functions, the pseudometric $\overline{\delta}^c$ is lower semi-continuous and hence belongs to $\cP$. We can then prove Theorem 2 of \cite{Chen25a}:

\begin{thm}
\label{thm:F-fixpoint}
For every $0 < c < 1$, the pseudometric  $\overline{\delta}^c$ is the least fixpoint of $\cF_c$ that is greater than, or equal to, the pseudometric $(x, y)\in E\times E \mapsto |obs(x) - obs(y)|\in [0,1]$.
\end{thm}
\noindent\textcolor{black}{\textbf{Return to the Brownian motion example:} Let $(B_t)_{t\geq0}$ be the standard Brownian motion on $\mathbb R$ starting from $0$. Recall that the usual topology $\mathcal O$ is generated by the $1$-bounded pseudometric $\Delta(x,y)=\min\{1,|x-y|\}$ for $x,y\in\mathbb R$. We claim that $\Delta$ is a fixed point for $\mathcal F_c$ in this case. Indeed, given any $x,y\in\mathbb R$ and $t\geq0$, if $\gamma$ is the coupling of $(P_t(x),P_t(y))$ given by $(x+B_t,y+B_t)$, then \[
W(\Delta)(P_t(x),P_t(y))\leq \mathbb E[\Delta(x+B_t,y+B_t)]=\Delta(x,y),
\]
which implies $\mathcal F_c(\Delta)\leq\Delta$; meanwhile, by taking $t=0$ in the supremum concerned in the definition of $\mathcal F_c$, we immediately have $\mathcal F_c(\Delta)\geq\Delta$. Hence, $\Delta$ is a fixed point of $\mathcal F_c$.} 

\textcolor{black}{In particular, whenever the observable function $obs$ is $1$-Lipschitz continuous under $\Delta$ (e.g., $obs(x)=\max\{0,1-|x|\}$ for $x\in\mathbb R$ is one such observable function), Theorem \ref{thm:F-fixpoint} applies and gives $\delta_0^c\leq\overline \delta^c\leq \Delta$.}
%$\overline{\delta}^c$ is a fixpoint of the functional $\cF_c$.

   \subsection{Through a real-valued logic}

Finally, similarly to \cite{Desharnais99b,Desharnais04,vanBreugel05a}, we show that the pseudometric $\overline{\delta}^c$ is characterized by a real-valued logic $\grammar$ which is defined inductively:
\begin{align*}
f \in \grammar & := q~|~ obs ~|~ \min\{ f_1, f_2 \} ~|~ 1-f ~|~ f \ominus q ~|~ \langle t \rangle f
\end{align*}
for all $f_1, f_2, f \in \grammar$, $q \in [0,1] \cap \mathbb{Q}$ and $t \in \mathbb{Q}_{ \geq 0}$. The interpretation can be found in \cite{Chen25a}. From this logic, we define the pseudometric $\lambda^c$:
\[\forall x,y\in E,\qquad \lambda^c(x, y) := \sup_{f \in \grammar} |f(x) - f(y)|.\]

Finally, we establish the equivalence between the logic characterization and the iterative definition of the pseudometric (Theorems 3 and 4 of \cite{Chen25a}):
\begin{thm}
\label{thm:F-logic}
 For every $0 < c < 1$, the pseudometric $\lambda^c$ is a fixpoint of $\cF_c$ and coincides with $\overline{\delta}^c$.
\end{thm}

\section{A trajectory pseudometric}
\label{sec:functional}

We are now ready to move on to trajectories. For this and next section, we will focus on the case when $\Omega$ is the space of continuous functions $\mathbb{R}_{\geq0}\to E$. We will start as in \cite{Chen25a} by defining a functional $\cG$ on the set $\cP$ of lower semi-continuous 1-bounded pseudometrics on the state space $E$. We show that the image of a continuous pseudometric is also continuous, which allows us to iteratively apply $\cG$ and to obtain a pseudometric $\overline{d}$ starting from $d_0(x,y) := |obs(x) - obs(y)|$. We can then show that $\overline{d}$ is a fixpoint of $\cG$. The key difference is that the cost function is no longer a pseudometric on the state space, but one on the trajectory space.

Recall that $\Omega$ is Polish (see Section \ref{sec:diffusions}). This is important in order to use the machinery of optimal transport. Similarly as in \cite{Chen25a}, we also need to add a discount factor $0 < c < 1$ and thus obtain a family of such functionals $\cG_c$ indexed by the discount factors and a corresponding family of pseudometrics obtained by iteratively applying $\cG_c$. 

%Furthermore, the case of $\cG$ is restricted to diffusion-like processes.

\subsection{Regularity of trajectory pseudometrics}

We will still use the three classes of pseudometrics $\cC \subseteq \cP \subseteq \cM$ defined in \cite{Chen25a} and reviewed in Section \ref{sec:previous-work}.

As we have stated beforehand, we will be working with trajectories in this section. We will therefore need to define a cost function on the space of trajectories $\Omega$. This is going to be done through the (discounted) uniform pseudometric:
\begin{defi}
\label{def:discounted-uniform-metric}
Given a pseudometric $m$ on $E$ and the discount factor $0< c\leq 1$, we define the \emph{discounted uniform pseudometric} $U_c(m)$ on the set of trajectories $\Omega$ as follows:
\[\forall \omega, \omega' \in \Omega, \qquad U_c(m) (\omega, \omega') := \sup_{t\geq 0} c^t m(\omega(t), \omega'(t)). \]
\end{defi}

\begin{rem}
If $m$ is a pseudometric on $E$,  then $U_c(m)$ is also a pseudometric on $\Omega$. Let us check quickly the triangular inequality: for $\omega_1, \omega_2$ and $\omega_3$ in $\Omega$,
\[ U_c(m) (\omega_1, \omega_3) = \sup_t c^t m(\omega_1(t), \omega_3(t)) \leq \sup_t c^t ( m(\omega_1(t), \omega_2(t)) + m(\omega_2(t), \omega_3(t))) \]
by triangular inequality for $m$. Since $c^t m(\omega_i (t), \omega_j(t)) \leq U_c(m)(\omega_i, \omega_j)$ for every $t$, we get the triangular inequality for $U_c(m)$ as desired.
\end{rem}

Note that if $c=1$, we get the uniform pseudometric
$U(m)$ with respect to $m$. We also have that
$U_c(m)\leq U(m)$ for every $0<c\leq1$.

In \cite{Chen25a}, we restricted the pseudometrics, which were used as cost functions, to lower semi-continuous ones ($\cP$) or to continuous ones ($\cC$).  When we are dealing with trajectories, our cost function is going to be a discounted uniform pseudometric. For that reason, we need to make sure that the regularity conditions on a pseudometric $m$ on the state space are also passed on to $U_c(m)$.

\begin{lem}
\label{lemma:m-lsc-Uc(m)-lsc}
For every $0<c\leq1$ and every pseudometric $m\in\cP$,
$U_c(m): \Omega\times\Omega\to\mathbb{R}$ is lower semi-continuous.
\end{lem}

\begin{proof}
For $r \in \mathbb R$, let us consider the set $U_c(m)^{-1}((r, \infty)) := \{ (\omega, \omega') ~|~ U_c(m)(\omega, \omega') > r \}$. We intend to show that this set is open in $\Omega \times \Omega$. We only need to consider the case when $r\in[0,1)$, since other values of $r$ would yield trivial $U_c(m)^{-1}((r,\infty))$ (the set is $\Omega\times\Omega$ for $r<0$, and $\emptyset$ for $r\geq 1$). 
%(which corresponds to the product topology generated by the metric $U(\Delta)$ on $\Omega$). 

Now fix $r\in[0,1)$. Consider $(\omega, \omega') \in U_c(m)^{-1}((r, \infty))$.  That is, $\sup_t c^t m (\omega(t), \omega'(t)) > r$, which implies that there exists $t$ such that $m (\omega(t), \omega'(t))> r c^{-t}$. Define the set $$O := \{ (x,y) ~|~ m(x,y) >rc^{-t} \} \subseteq E \times E.$$ This set $O$ is open in $E \times E$ since the function $m$ is lower semi-continuous. Since the pair $(\omega(t), \omega'(t))$ is in $O$, there exists $q > 0$ \footnote{Note that this $q$ also depends on how the topology on $E \times E$ is metrized, for instance $p$-distance or max distance} such that $B_q \times B'_q \subseteq O$ where the two sets $B_q$ and $B'_q$ are defined as open balls in $E$:
\[ B_q := \{ z ~|~ \Delta (z, \omega(t)) < q \} \qquad \text{and} \qquad B'_q := \{ z ~|~ \Delta (z, \omega'(t)) < q \}. \]
Further define two sets $A$ and $A'$ of trajectories as
%\[ A := \{ \theta ~|~ U(\Delta)(\omega, \theta) < q\} \qquad \text{and} \qquad A' := \{ \theta ~|~ U(\Delta)(\omega', \theta) < q\}. \]
\[ A := \{ \theta ~|~ \Delta(\omega(t), \theta(t)) < q\} \qquad \text{and} \qquad A' := \{ \theta ~|~ \Delta(\omega'(t), \theta(t)) < q\}. \]

These are open sets in $\Omega$ and thus $A \times A'$ is open in $\Omega \times \Omega$. 

We claim that $A \times A'\subseteq U_c(m)^{-1}((r, \infty))$. To see this, consider $\theta \in A$ and $\theta' \in A'$. This means that $\Delta (\omega(t), \theta(t)) < q$, i.e. $\theta(t) \in B_q$. Similarly, $\theta'(t) \in B'_q$. Thus, $(\theta(t), \theta'(t)) \in B_q \times B'_q \subseteq O$, which means that $c^t m(\theta(t), \theta'(t)) >r $. Hence, $(\theta,\theta')\in U_c(m)^{-1}((r, \infty))$, which proves the claim.

Finally, since $\omega\in A$ and $\omega' \in A'$, the set $A \times A'$ is an open neighbourhood of $(\omega, \omega')$ in $U_c(m)^{-1}((r, \infty))$. We conclude that $U_c(m)^{-1}((r, \infty))$ is open. 
\end{proof}

The definitions of $U_c$ and $\mathcal G_c$ below, as well
as the preceding lower semi-continuity result, include
the case $c=1$. Continuity of $U_c(m)$ for continuous $m$
requires strict discounting in general. Accordingly,
the subsequent continuity results and iterative
construction are established for $0<c<1$.

\begin{lem}
\label{lemma:m-cont-Uc(m)-cont}
For every $0<c<1$ and every pseudometric $m \in \cC$, $U_c(m):\Omega \times \Omega \to \mathbb{R}$ is continuous.
\end{lem}

\begin{proof}
It is enough to show that if $(\omega_n)_{n \in \mathbb{N}}$ is a sequence of trajectories that converges to $\omega$ in the topology of $\Omega$, then $\lim_{n \to \infty} U_c(m)(\omega, \omega_n) = 0$.

Assume it is not the case. This means that there exists $\epsilon > 0$ and an increasing sequence of natural numbers $(n_k)_{k \in \mathbb{N}}$ such that $U_c(m) (\omega, \omega_{n_k}) \geq 3 \epsilon$ for all $k\in\mathbb{N}$. This means that for every $k$, there exists a time $t_k$ such that $c^{t_k} m (\omega(t_k), \omega_{n_k}(t_k)) \geq 2 \epsilon$. Since $m$ is 1-bounded, those times $t_k$ are bounded by $T:=\ln (2 \epsilon)/ \ln c$.

For that reason we may assume that  the sequence $(t_k)_{k \in \mathbb{N}}$ converges to some time $t$ (otherwise there is a subsequence that converges, so we can consider that subsequence instead and the corresponding trajectories $\omega_{n_k}$). We have the following inequalities for every $k$:
\begin{align*}
2 \epsilon
& \leq c^{t_k} m(\omega (t_k), \omega_{n_k}(t_k))\\
& \leq c^{t_k} m(\omega (t_k), \omega(t)) + c^{t_k} m(\omega (t), \omega_{n_k}(t_k)).
\end{align*}
Since $m$ is continuous and so are the trajectories,  $\lim_{k \to \infty} m(\omega (t_k), \omega(t)) = 0$. This means that there exists $K$ such that for all $k \geq K$,  $m(\omega (t_k), \omega(t)) < \epsilon$. Since $c< 1$, this implies that $c^{t_k} m(\omega (t_k), \omega(t)) < \epsilon$ and thus using previous inequality,  $\epsilon <   c^{ t_k}  m(\omega(t), \omega_{n_k}(t_k))$. Pushing this further, using that $c< 1$ once more, we get that for all $k \geq K$,  $\epsilon < m(\omega(t), \omega_{n_k}(t_k))$.

Now recall that $(\omega_n)_{n \in \mathbb{N}}$ is a sequence of trajectories that converges to $\omega$ (in the topology of $\Omega$), which means that  $\lim_{n \to \infty} \sup_{s\in[0,T]}\Delta(\omega_n(s), \omega(s)) = 0$. Using the triangular inequality, we have that
\[ \Delta (\omega_{n_k}(t_k), \omega(t))
\leq \Delta (\omega_{n_k}(t_k), \omega(t_k)) + \Delta (\omega(t_k), \omega(t))
\leq \sup_{s\in[0,T]}\Delta(\omega_{n_k}(s), \omega(s)) + \Delta (\omega(t_k), \omega(t)). \]
The second term of the right-hand side converges to 0 as $k \to \infty$ because $\omega$ is continuous.
This means that $\lim_{k \to \infty} \Delta (\omega_{n_k}(t_k), \omega(t)) = 0$, 
%Using Lemma \ref{lemma:subtopologies}, 
and hence $\lim_{k \to \infty} m (\omega_{n_k}(t_k), \omega(t)) = 0$ which directly contradicts that $\epsilon < m(\omega(t), \omega_{n_k}(t_k))$ for every $k \geq K$.
\end{proof}

\subsection{The family of functionals}

%we defined in Section \ref{sec:gen-CT} compared the distributions on the state space $P_t(x)$. 
%The new functional that we will define here compares  the probability distributions $\mathbb{P}^x$ on the space of trajectories through transport theory.  Similarly to that in Section \ref{sec:gen-CT}, this functional is defined for diffusions as we need the map $x \mapsto \mathbb{P}^x$  to be continuous.

%Just as the first functional $\cF_c$ is defined with a discount factor $c$, the second functional that we will define also involves a discount factor $c$, it is denoted $\cG_c$.  The functional $\cG_c$ can be defined for $0< c \leq 1$ but we will later require $0<c<1$. The transport cost between two trajectories is given by the discounted uniform metric.

Given a discount factor $0 < c\leq 1$,  we define the functional $\cG_c: \cP \to \cM$ as follows: for every pseudometric $m \in \cP$ and every two states $x,y$,
\[ \cG_c(m)(x,y) := W(U_c(m))(\mathbb{P}^x, \mathbb{P}^y).\]
This is well-defined as we have proven in Lemma \ref{lemma:m-lsc-Uc(m)-lsc} that if $m$ is lower semi-continuous, then so is $U_c(m)$. Using the Kantorovich duality, we have:
\begin{align*}
\cG_c(m)(x,y)  & =  \min_{\gamma \in \Gamma(\mathbb{P}^x, \mathbb{P}^y)} \int U_c(m) ~ \dee \gamma\\
& =  \sup_{h \in \cH(U_c(m))} \left(\int h ~ \dee  \mathbb{P}^x - \int h ~ \dee  \mathbb{P}^y\right).
\end{align*}
The discussion before Lemma \ref{lemma:wasserstein-limit} ensures that $\cG_c(m)$ is indeed in $\cM$.

All the precautions that we had to take in our previous work still apply here: for the iterative construction, we need to restrict to the class $\cC$ because the functional $\cG_c$ maps $\cC$ into itself, as we will establish below. However, given an increasing sequence of pseudometrics in $\cC$, the supremum of the sequence belongs to $\cP$ in general, and may not be in $\cC$.  Therefore, the Knaster-Tarski theorem may not be applicable, and hence we need to go through similar steps as for $\cF_c$ to establish the acquired 
supremum as a fixpoint of $\cG_c$.\\

As a direct consequence from the definition of $\cG_c$, we have that, as a functional on $\cP$, $\cG_c$ is monotonic (for every $0 < c \leq 1$): if $m_1 \leq m_2$ in $\cP$, then $\cG_c(m_1) \leq \cG_c(m_2)$.

\begin{lem}
\label{lemma:order-m-Fc-Gc}
For every $0<c\leq1$, every pseudometric $m\in\cP$, and every pair of states~$x,y$,
\[ m(x,y) \leq \cF_c(m)(x,y) \leq \cG_c(m)(x,y). \]
\end{lem}

\begin{proof}
The first inequality is just a direct consequence of the definition of $\cF_c$ and can be found in \cite{Chen25a}. For the second inequality: pick $t \geq 0$ and $h: E \to [0,1] \in \cH(m)$. Define $h_t : \Omega \to [0,1]$ as $h_t(\omega) := c^t h(\omega(t))$. Then note that for every pair of trajectories $\omega, \omega'$,
\begin{align*}
|h_t(\omega) - h_t(\omega')|
& = c^t |h(\omega(t)) - h(\omega'(t))|\\
& \leq c^t m(\omega(t), \omega'(t))\\
& \leq U_c(m) (\omega, \omega').
\end{align*}
This means that $h_t \in \cH(U_c(m))$ and thus
\begin{align*}
\cG_c(m)(x,y)
& \geq \int h_t ~ \dee \mathbb{P}^x - \int h_t ~ \dee \mathbb{P}^y\\
& = c^t \left(\int h ~ \dee P_t(x) - \int h ~ \dee P_t(y) \right)\\
& = c^t \left( \hat{P}_t h(x) - \hat{P}_t h(y) \right).
\end{align*}
Since $t\in\mathbb{R}_{\geq0}$ is arbitrary, we conclude our proof.
\end{proof}

\begin{cor}
\label{cor:fixpoint-G-implies-F}
For every $0<c\leq1$, a fixpoint for $\cG_c$ is also a fixpoint for $\cF_c$.
\end{cor}

The proof of this corollary directly follows from the observation that for a pseudometric $m \in \cP$ that is a fixpoint of $\cG_c$ and two states $x$ and $y$:
\[ m(x, y) \leq \cF_c(m) (x,y) \leq \cG_c(m)(x,y) = m(x,y). \]

As we have stated, a key argument in our construction is that $\cG_c: \cC \to \cC$ when $0<c<1$.

\begin{lem}
\label{lemma:Gc-cont-is-cont}
For every $0<c<1$ and every pseudometric $m \in \cC$, the topology on $E$ generated by $\cG_c(m)$ is a subtopology of the original topology $\cO$.
\end{lem}

\begin{proof}
Using Lemma \ref{lemma:m-cont-Uc(m)-cont}, we know that $U_c(m)$ is continuous with respect to the topology on $\Omega \times \Omega$ for every $0<c<1$.

In order to show that $\cG_c(m)$ generates a topology on $E$ which is a subtopology of $\cO$ (generated by $\Delta$), it is enough to show that for a fixed state $x \in E$, the map $y \mapsto \cG_c(m)(x,y)$ is continuous on $E$. To this end, we pick a sequence $(y_n)_{n \in \mathbb{N}}$ in $E$ such that $y_n \underset{n \to \infty}{\longrightarrow} y$ in the topology $\cO$ with $y \in E$. Recall that $\mathbb{P}^{y_n}$ converges weakly to $\mathbb{P}^y$. Then,
\begin{align*}
 \lim_{n \to \infty} \cG_c(m)(x, y_n)
 & =  \lim_{n \to \infty} W(U_c(m))(\mathbb{P}^x, \mathbb{P}^{y_n}) \\
& = \lim_{n \to \infty }  \min \left\{ \left.  \int U_c(m) \dee \gamma ~ \right| ~ \gamma \in \Gamma (\mathbb{P}^x, \mathbb{P}^{y_n}) \right\}.
\end{align*}
Using Theorem 5.20 of \cite{Villani08}, we know that the optimal transport plan $\pi_n$ between $\mathbb{P}^x$ and $\mathbb{P}^{y_n}$ converges, when the cost function is the continuous function $U_c(m)$, up to extraction of a subsequence, to an optimal transport plan $\pi$ for $\mathbb{P}^x$ and $\mathbb{P}^y$. This means that $\cG_c(m)(x, y_n)$ converges  up to extraction of a subsequence, to $\cG_c(m)(x,y)$.

However, the whole sequence $(\cG_c(m)(x, y_n))_{n \in \mathbb{N}}$ converges to $\cG_c(m)(x,y)$ for the following reason. Assuming otherwise, since $0 \leq \cG_c(m)(x, y_n) \leq 1$ for all $n \in \mathbb{N}$, there exists another subsequence $(z_k)_{k \in \mathbb{N}}$ such that $( \cG_c(m)(x, z_k))_{k \in \mathbb{N}}$ converges to a limit $l \neq \cG_c(m)(x, y)$. We also have that $\lim_{k \to \infty} z_k = y$, and hence by Theorem 5.20 of \cite{Villani08}, there is a subsequence $(z_{k_j})_{j \in \mathbb{N}}$ such that $\lim_{j \to \infty}\cG_c(m)(x, z_{k_j}) = \cG_c(m)(x, y)$ which yields a contradiction.
\end{proof}

\subsection{The fixpoint pseudometric}

As mentioned above, we will construct an increasing sequence of continuous
pseudometrics and prove that the
supremum of this sequence is a fixpoint
of $\cG_c$. 

Lemma \ref{lemma:Gc-cont-is-cont} enables us to define for every $0 < c< 1$ the following sequence of pseudometrics on $E$: for every $x,y\in E$,
\begin{align*}
d_0^c(x,y) & := |obs(x) - obs(y)|,\\
d_{n+1}^c(x,y) & := \cG_c (d_n^c)(x,y)\;\textup{ for }n\geq0.
\end{align*}
Then, we can further define $\overline{d}^c := \sup_{n \in \mathbb{N}} d_n^c$ (which is also a limit since the sequence is non-decreasing by Lemma \ref{lemma:order-m-Fc-Gc}). As a supremum of continuous functions,  the function $\overline{d}^c$ is lower semi-continuous and thus belongs to $\cP$ for every $0 < c < 1$.\\

We are now ready to prove that $\overline{d}^c$ is a fixpoint of $\cG_c$. Since $\cG_c(\overline{d}^c)$ is defined as the optimal transport cost when the transport function is $U_c(\overline{d}^c)$, in order to study $\cG_c(\overline{d}^c)$, we need to relate $U_c(\overline{d}^c)$ to $U_c(d_n^c)$ which the next lemma does.

\begin{lem}
\label{lemma:Ud-bar-is-sup-Ud-n}
For every pair of  trajectories $\omega, \omega'\in\Omega$, 
\[ U_c(\overline{d}^c)(\omega, \omega') =  \lim_{n \to \infty} U_c(d_n^c) (\omega, \omega') = \sup_{n\in\mathbb{N}} U_c(d_n^c) (\omega, \omega').  \]
\end{lem}

\begin{proof}
We omit the superscript $c$ in all the pseudometrics $d_n$ and $\overline d$ throughout this proof. 

First note that $U_c(d_{n+1}) \geq U_c(d_n)$ since $U_c$ is monotonic, which shows the second equality above. Further,
\begin{align*}
U_c(\overline{d})(\omega, \omega')
&= \sup_{t \geq 0} c^t\, \overline{d} (\omega(t), \omega'(t)) \\
&= \sup_{t \geq 0} c^t \sup_{n\in\mathbb{N}} d_n (\omega(t), \omega'(t)) \\
& = \sup_{n\in\mathbb{N}} \sup_{t \geq 0}  c^t d_n (\omega(t), \omega'(t)) \\
& = \sup_{n\in\mathbb{N}} U_c(d_n) (\omega, \omega').
\end{align*}
\end{proof}

\begin{thm}
\label{thm:m-fixpointG}
The pseudometric $\overline{d}^c$ is a fixpoint of $\cG_c$.
\end{thm}

\begin{proof}
Again, we omit the superscript $c$ in $d_n$ and $\overline{d}$ throughout this proof. 

According to Lemma \ref{lemma:m-cont-Uc(m)-cont} and Lemma \ref{lemma:Gc-cont-is-cont}, $\{U_c(d_n):n\geq1\}$ forms an increasing sequence of continuous pseudometrics and converges to $U_c(\overline d)$ which is a lower semi-continuous pseudometric. Thus, by Lemma \ref{lemma:wasserstein-limit},
for every $x,y\in E$,
\[
\cG_c(\overline d)(x,y)=W(U_c(\overline d))(\mathbb{P}^x,\mathbb{P}^y)=\lim_{n\to\infty}W(U_c(d_n))(\mathbb{P}^x,\mathbb{P}^y)=\lim_{n\to\infty}\cG_c(d_n)(x,y)=\overline d(x,y).
\]

\end{proof}

The following lemma is similar to Lemma 6 of \cite{Chen25a} and the proof is essentially the same.

\begin{lem}
\label{lemma:comparison-fixpointG}
For every $0 < c < 1$, if $m$ is a pseudometric in $\cP$ such that $m$ is a fixpoint for $\cG_c$, and $m(x,y) \geq |obs(x) - obs(y)|$ for every $x,y\in E$,
then $m \geq \overline{d}^c$.
\end{lem}

This lemma leads to the following important result, which is the analogue to Theorem \ref{thm:F-fixpoint} in the trajectory framework:
\begin{thm}\label{thm:G-fixpoint}
For every $0<c<1$, the pseudometric $\overline{d}^c$ is the least fixpoint of $\cG_c$ that is greater than, or equal to, the pseudometric $(x, y)\in E\times E \mapsto |obs(x) - obs(y)|\in[0,1]$.
\end{thm}

\noindent\textbf{Return to the Brownian motion example, again:} Let us examine our Brownian motion example under this trajectory pseudometric setting. Once again, consider the standard Brownian motion $(B_t)_{t\geq0}$ on $\mathbb R$ (in which case $\Omega=C(\mathbb R_{\geq0};\mathbb R)$, the set of continuous functions $R_{\geq0} \to \mathbb R$), the $1$-bounded pseudometric $\Delta(x,y)=\min\{1,|x-y|\}$ for $x,y\in\mathbb R$, and the observable function $obs(x)=\max\{0,1-|x|\}$. Given $x,y\in\mathbb R$, take $\gamma$ to be the coupling of $\mathbb P^x$ and $\mathbb P^y$ given by $((x+B_t)_{t\geq0},(y+B_t)_{t\geq0})$, meaning that
\textcolor{black}{
\[ \gamma(B \times B') = \mathbb{P}^0 \left(\{ \omega ~|~ \omega + x \in B ~\text{and}~ \omega + y \in B'  \} \right), \]
where $\omega + z$ is the trajectory $\omega$ translated by $z$.}
Note that $\Delta$ is translation invariant, and hence for every $(\omega,\omega')\in\Omega\times\Omega$ sampled under $\gamma$,
\[
U_c(\Delta)(\omega,\omega')=\sup_{t\geq0}c^t\Delta(\omega(t),\omega'(t))=\sup_{t\geq0}c^t\Delta(x+B_t,y+B_t)=\Delta(x,y).
\]
Thus, \[
\mathcal G_c(\Delta)(x,y)\leq\int U_c(\Delta)\dee \gamma\leq\Delta(x,y).\]
Combining this with Lemma \ref{lemma:order-m-Fc-Gc}, we conclude that $\mathcal G_c(\Delta)=\Delta$, i.e., $\Delta$ is a fixed point for $\mathcal G_c$.

\textcolor{black}{In particular, whenever the observable function $obs$ is $1$-Lipschitz continuous under $\Delta$ (e.g., $obs(x)=\max\{0,1-|x|\}$ for $x\in\mathbb R$ is one such observable function), Theorem \ref{thm:G-fixpoint} applies and gives $d_0^c\leq\overline d^c\leq \Delta$.}

\section{The trajectory pseudometric defined through a real-valued logic}
\label{sec:logicG}

Similarly to \cite{Desharnais99b,Desharnais04,vanBreugel05a} for discrete-time studies and to our previous work, the pseudometric $\overline{d}^c$ obtained from the functional $\cG_c$ can also be described through a real-valued logic. 
\subsection{The real-valued logic} Clearly, the desired logic needs to handle both the states and the trajectories, so it will contain two parts: $\cL_\sigma$ and $\cL_\tau$.
\subsubsection{Definition of the logic:}
We define the logic inductively in two parts as follows:
\begin{align*}
f \in \cL_\sigma & := q~|~ obs ~|~ 1-f ~|~ \int g,\\
g \in \cL_\tau & := f \circ ev_t ~|~  \min \{ g_1, g_2\}  ~|~ \max \{ g_1, g_2\} ~|~ g \ominus q ~|~ g \oplus q,
\end{align*}
where $q \in \mathbb{Q} \cap [0,1]$, $f \in \cL_\sigma$, $g, g_1, g_2 \in \cL_\tau$ and $t \in \mathbb{Q}_{\geq 0}$. 

\subsubsection{Interpretation of the logic:}
For a fixed discount factor $0< c< 1$, the expressions in $\cL_\sigma$ are interpreted as functions $E \to [0,1]$, and the expressions in $\cL_\tau$ are interpreted as functions $\Omega \to [0,1]$. More specifically, the terms of $\cL_\sigma$ are interpreted as the corresponding ones in $\grammar$:  for a state $x \in E$,
\begin{align*}
q(x) & = q, \\
obs(x) & = obs(x), \\
(1-f)(x) & = 1 - f(x), \\
\left(\int g\right)(x) & = \int g ~ \dee \mathbb{P}^x.
\end{align*}
The terms of  $\cL_\tau$ are interpreted for a trajectory $\omega\in\Omega$ as:
\begin{align*}
( f \circ ev_t) (\omega) & = c^t f(\omega(t)),\\
( \min \{g_1, g_2\}) (\omega)& = \min \{g_1(\omega), g_2(\omega)\},\\
( \max \{g_1, g_2\}) (\omega)& = \max \{g_1(\omega), g_2(\omega)\},\\
(g \ominus q )(\omega) & = \max \{ 0, g(\omega) - q \}, \\
(g \oplus q )(\omega) & = \min \{ 1, g(\omega) + q \}.
\end{align*}

Whenever we want to emphasize the fact that the expressions are interpreted for a certain discount factor $0<c<1$, we will write $\cL^c_\sigma$ and $\cL^c_\tau$.
\subsubsection{Additional useful expressions: }
From $\cL_\tau$, we can define additional expressions in $\cL_\sigma$ as follows:
\begin{align*}
f \ominus q & = \int [(f \circ ev_0) \ominus q],\\
\min \{ f_1, f_2\} & = \int \min \{ f_1 \circ ev_0, f_2 \circ ev_0 \},\\
\max\{ f_1, f_2 \} & = 1 - \min \{1-f_1, 1-f_2 \},\\
f \oplus q &  = 1 - ((1 - f) \ominus q).
\end{align*}
Their interpretations as functions $E \to [0,1]$ are the same as for $\grammar$.

\subsection{Some properties of the logic} Let us present some technical results on the logic defined above. They will be useful later in establishing the relation between the logic and the pseudometric $\overline{d}^c$. We start with the continuity property of the expressions in the logic.
\begin{lem}
\label{lemma:logic-st-continuous}
For every expression $f \in \cL_\sigma$, the function $x \mapsto f(x)$ is continuous, and for every expression $g \in \cL_\tau$, the function $\omega \mapsto g(\omega)$ is continuous.
\end{lem}

\begin{proof}
This is done by induction on the structure of $f$ and $g$. There are two cases that are not straightforward.
\begin{itemize}
\item First, if $f = \int g$ with $g \in \cL_\tau$, then, since $x \mapsto \mathbb{P}^x$ is continuous and $g$ is continuous in $\Omega$, we get that $x \mapsto \int g ~ \dee \mathbb{P}^x$ is continuous in $E$.
\item Second, if $g =  f \circ ev_t$ with $f \in \cL_\sigma$ and $t \in \mathbb{Q}_{\geq 0}$, then since trajectories are continuous and $f$ is continuous on the state space, we get the desired result.
\end{itemize}\end{proof}

Next, we will prove that certain functions on the trajectories can be approximated by functions in the logic $\cL_\tau$. This is done in several steps. We first state an approximation result on a given pair of trajectories.

\begin{lem}
\label{lemma:prep-magie}
Consider a function $h: \Omega \to [0,1]$ and two trajectories $\omega, \omega'\in \Omega$ such that there exists $f$ in the logic $\cL_\sigma$ and a time $s \in \mathbb{Q}_{\geq 0}$ such that $$|h(\omega) - h(\omega')| \leq c^s|f(\omega(s)) - f(\omega'(s))|.$$ Then, for every $\delta > 0$, there exists $g$ in the logic $\cL_\tau$ such that 
\[
|h(\omega) -g(\omega)| \leq 2 \delta\;\textit{ and }\;|h(\omega') - g(\omega')| \leq 2 \delta.
\]
\end{lem}

\begin{proof}
Without loss of generality, we may assume $h(\omega) \geq h(\omega')$ and $f(\omega(s)) \geq f(\omega'(s))$ (otherwise consider $1-f$ instead of $f$).

Pick $p,q,r \in \mathbb{Q} \cap [0,1]$ such that
\begin{align*}
p & \in [c^s f(\omega'(s)) - \delta, c^s f(\omega'(s))],\\
q & \in [h(\omega) - h(\omega') - \delta, h(\omega) - h(\omega')],\\
r & \in [ h(\omega'), h(\omega') + \delta].
\end{align*}
Define $g := ( \min \{ (f \circ ev_s) \ominus p, q \} ) \oplus r$. Let us evaluate $g$ on $\omega$ and $\omega'$. 

First, by the range of $p$ and the assumption $f(\omega(s))\geq f(\omega'(s))$, we have that
\begin{align*}
(f \circ ev_s)(\omega) & = c^s f(\omega(s)),\\
((f \circ ev_s) \ominus p)(\omega) & = \max \{ 0, c^s f(\omega(s)) - p\} \\
& \in [c^s [f(\omega(s)) - f(\omega'(s))], c^s [f(\omega(s)) - f(\omega'(s))]+ \delta].
\end{align*}
Since $q \leq  h(\omega) - h(\omega')\leq c^s [f(\omega(s)) - f(\omega'(s))]$, we further get 
\[
(\min \{ (f \circ ev_s) \ominus p, q \})(\omega)
= q.
\]
%\begin{align*}
%   (\min \{ (f \circ ev_s) \ominus p, q \})(\omega)
%& = q\\
%q + r& \in [h(\omega) - \delta, h(\omega) + \delta],\\
%g(\omega) & = \min \{ 1, q + r\} \in [h(\omega) - \delta,  h(\omega) + \delta] ~~ \text{as } h(\omega) \leq 1.
%\end{align*}
Finally, by the definition of $g$ and the range of $q,r$, we conclude 
\[
g(\omega) = \min \{ 1, q + r\} \in [h(\omega) - \delta,  h(\omega) + \delta] ~~ \text{as } h(\omega) \leq 1.
\]

We can conduct a similar derivation for $g(\omega')$ as:
\begin{align*}
(f \circ ev_s)(\omega') & = c^s f(\omega'(s)),\\
((f \circ ev_s) \ominus p)(\omega')
& = \max \{ 0, c^s f(\omega'(s)) - p\} \in [0, \delta],\\
(\min \{ (f \circ ev_s) \ominus p, q \})(\omega')
& \in [0, \delta],\\
g(\omega') & = \min \{ 1,  (\min \{ (f \circ ev_s) \ominus p, q \})(\omega') + r\}\\
& \in [h(\omega'), h(\omega') + 2 \delta].
\end{align*}
\end{proof}

The following result  from \cite{Ash72} (Lemma A.7.2) is a key ingredient in our approximation theory. It is also used in~\cite{vanBreugel05a} for discrete-time processes.
\begin{lem}
\label{lemma:magie}
Let $X$ be a compact Hausdorff space.  Let $A$ be a subset of the set of continuous functions $X \to \mathbb{R}$ such that if $f, g \in A$, then $\max\{ f,g\}$ and $\min \{f,g\}$ are also in $A$.  Consider a continuous function $h$ that can be approximated at each pair of points by functions in $A$, meaning that 
\[ \forall x, y \in X~ \forall \epsilon > 0~ \exists g \in A ~ |h(x) - g(x)| \leq \epsilon \text{ and } |h(y) - g(y)| \leq \epsilon \]
Then, $h$ can be approximated by functions in $A$, meaning that $$\forall \epsilon > 0 ~ \exists g \in A ~ \forall x \in X~ |h(x) - g(x)| \leq \epsilon.$$
\end{lem}

%We will later on use Lemma \ref{lemma:magie} to show that a function on trajectories in $\cH(U_c(\ell^c))$ can be approximated by functions in $\cL_\tau$. 
Combining the three lemmas above, we obtain the following approximation result. 
\begin{cor}\label{cor:approx by functions in logic}
Consider a continuous function $h: \Omega \to [0,1]$ such that for every two trajectories $\omega, \omega'\in\Omega$, there exists $f$ in the logic $\cL_\sigma$ and a time $s \in \mathbb{Q}_{\geq 0}$ such that $$|h(\omega) - h(\omega')| \leq c^s |f(\omega(s)) - f(\omega'(s))|.$$ Then, for every compact subset $\Omega'\subseteq\Omega$, the function $h$ can be approximated by functions (that are interpretations of expressions) in $\cL_\tau$ on $\Omega'$.
\end{cor}

\begin{proof}
We have proven in Lemma \ref{lemma:prep-magie} that such a function $h$ can be approximated at pairs of trajectories by functions from $\cL_\tau$ and $s \in \mathbb{Q}_{\geq 0}$. By Lemma \ref{lemma:logic-st-continuous}, all the functions in $\cL_\tau$ are continuous. Since $\cL_\tau$ is closed under $\min$ and $\max$, we can apply Lemma \ref{lemma:magie} to conclude that, when restricted on any compact set of trajectories, the function $h$ can be approximated by functions in $\cL_\tau$.
\end{proof}

\subsection{The pseudometric arising from the logic}

The pseudometric we derive from the logic $\cL_\sigma$ corresponds to how different the test results are when the process starts from $x$  compared to when it starts from $y$.

Given a fixed discount factor $0< c<1$, we can define the pseudometric $\ell^c$:
\[\forall x,y\in E,\quad\ell^c (x,y) := \sup_{f \in \cL^c_\sigma} |f(x) - f(y)| = \sup_{f \in \cL^c_\sigma} (f(x) - f(y)).\]
The last equality holds because for every $f \in \cL^c_\sigma$, $1-f$ is also in $\cL^c_\sigma$.

We can then follow a similar proof to the one of Theorem \ref{thm:F-logic} to establish that $\ell^c$ is a fixpoint of $\cG_c$.

\begin{thm}
\label{thm:mst-fixpoint}
The pseudometric $\ell^c$ is a fixpoint of $\cG_c$: $\ell^c = \cG_c(\ell^c)$.
\end{thm}

\begin{proof}
We will omit writing the index $c$ in this proof.

We already know that $\cG(\ell) \geq \ell$ (Lemma \ref{lemma:order-m-Fc-Gc}), so we only need to prove the other inequality. Pick two states $x,y\in E$ and recall that $\cG(\ell) (x,y) = W(U(\ell))(\mathbb{P}^x, \mathbb{P}^y)$ where
\[ \forall\omega,\omega'\in\Omega,\quad U(\ell)(\omega, \omega') = \sup_{t \geq 0, f \in \cL_\sigma} c^t| f(\omega(t)) - f(\omega'(t)) |. \]
Since for every $f \in \cL_\sigma$ the map $z \mapsto f(z)$ is continuous (Lemma \ref{lemma:logic-st-continuous}) and the trajectories are continuous, it is enough to consider rational times only and we have that
\[ U(\ell)(\omega, \omega') = \sup_{t \in \mathbb{Q}_{\geq 0}, f \in \cL_\sigma} c^t | f(\omega(t)) - f(\omega'(t)) |. \]
Now, both the logic $\cL_\sigma$ and the set of non-negative rationals $\mathbb{Q}_{\geq 0}$ are countable, so we can enumerate their elements as $\cL_{\sigma} = (f_k)_{k \in \mathbb{N}}$ and $\mathbb{Q}_{\geq 0} = (s_k)_{k \in \mathbb{N}}$. Define the map
\[ c_k(\omega, \omega') = \max_{ i,j = 1, ..., k} c^{s_i} |f_j(\omega(s_i)) - f_j(\omega'(s_i))| .\]
Note that $\lim_{k \to \infty} c_k (\omega, \omega') = U(\ell)(\omega, \omega')$ for every two trajectories $\omega, \omega'$. By Lemma \ref{lemma:wasserstein-limit},
%\footnote{The set $\Omega$ equipped with the Skorohod topology is Polish. Since all the trajectories are continuous, the uniform topology is equivalent to the Skorohod topology }
we get that $\lim_{k \to \infty} W(c_k) (\mathbb{P}^x, \mathbb{P}^y) = W(U(\ell))(\mathbb{P}^x, \mathbb{P}^y)$. 

It is thus enough to prove that $W(c_k) (\mathbb{P}^x, \mathbb{P}^y)\leq\ell(x,y)$ for every $k\in\mathbb{N}$. To this end, fix $k \in \mathbb{N}$. There exists $h$ such that \[
\forall \omega, \omega'\in\Omega,~ |h(\omega) - h(\omega')| \leq c_k(\omega, \omega')
\]
and 
\[ W(c_k) (\mathbb{P}^x, \mathbb{P}^y) =\int h~ \dee \mathbb{P}^x - \int h~ \dee \mathbb{P}^y. \]
Using Lemma \ref{lemma:logic-st-continuous}, we know that $c_k$ is continuous (as a maximum of finitely many continuous functions), which means that $h$ is also continuous. By the first condition about $h$ and the definition of $c_k$, we know that for every pair of trajectories $\omega, \omega'\in\Omega$, there exist $i,j \leq k$ such that $$|h(\omega) - h(\omega')| \leq c^{s_i} |f_j(\omega(s_i)) - f_j(\omega'(s_i))|.$$ 

Since $\mathbb{P}^x$ and $\mathbb{P}^y$
are tight ($\Omega$ is Polish), for every $\epsilon>0$, there exists compact subset $\Omega'\subseteq\Omega$ such that $\mathbb{P}^x(\Omega\setminus\Omega')\leq\frac{\epsilon}{4}$ and $\mathbb{P}^y(\Omega\setminus\Omega')\leq\frac{\epsilon}{4}$. Then, by Corollary \ref{cor:approx by functions in logic}, there exists a sequence $(g_n)_{n\in\mathbb{N}}$ in $\cL_\tau$ that converges to $h$ on $\Omega'$. In particular, when $n$ is sufficiently large, \[
\left\vert\int_{\Omega'}g_n\,\dee\mathbb{P}^x-\int_{\Omega'}h\,\dee\mathbb{P}^x\right\vert \leq\frac{\epsilon}{4}
\]
and similarly for $\mathbb{P}^y$, and hence
\[
\left\vert\int_{\Omega}g_n\,\dee\mathbb{P}^x-\int_{\Omega}h\,\dee\mathbb{P}^x\right\vert \leq\left\vert\int_{\Omega'}g_n\,\dee\mathbb{P}^x-\int_{\Omega'}h\,\dee\mathbb{P}^x\right\vert+\int_{\Omega\setminus\Omega'}|g_n-h|\,\dee\mathbb{P}^x\leq\frac{\epsilon}{2},
\]
and similarly for $\mathbb{P}^y$. We can thus apply the triangle inequality and get that 
\begin{align*}
\int h~ \dee \mathbb{P}^x - \int h~ \dee \mathbb{P}^y 
& \leq\left\vert\int_{\Omega}g_n\,\dee\mathbb{P}^x-\int_{\Omega}g_n\,\dee\mathbb{P}^y\right\vert+\epsilon\\
&=\left\vert\left(\int g_n\right)(x)-\left(\int g_n\right)(y)\right\vert+\epsilon\\
& \leq \sup_{f \in \cL_\sigma} |f(x) - f(y)|+\epsilon\\
&= \ell(x,y)+\epsilon.
\end{align*}
Since $\epsilon>0$ is arbitrary, we conclude that $W(c_k) (\mathbb{P}^x, \mathbb{P}^y)\leq\ell(x,y)$.
%and we thus know that $h$ can be approximated by functions that are interpretations of expressions in $\cL_\tau$.
%This means in particular that
%\begin{align*}
%\int h~ \dee \mathbb{P}^x - \int h~ \dee \mathbb{P}^y 
%& \leq \sup_{g \in \cL_\tau} \left(\int g~ \dee \mathbb{P}^x -  \int g~ \dee \mathbb{P}^y\right) \\
%& =  \sup_{g \in \cL_\tau}\left[ \left( \int g \right) (x) - \left( \int g \right)(y)\right] \\
%& \leq \sup_{f \in \cL_\sigma} |f(x) - f(y)| = \ell(x,y).
%\end{align*}
%Moving back to $W(U(\ell))$, we thus get that 
%\[ W(U(\ell))(\mathbb{P}^x, \mathbb{P}^y) = \lim_{k \to \infty} W(c_k) (\mathbb{P}^x, \mathbb{P}^y) \leq \ell(x,y) \]
%which concludes the proof.
\end{proof}

\subsection{Equality of the pseudometrics $\overline{d}^c$ and $\ell^c$}

In Theorem \ref{thm:F-logic}, we proved that the pseudometric $\lambda^c$ arising from the real-valued logic $\grammar$ was identical to $\overline{\delta}^c$ generated by the functional $\cF_c$. Such a relation also holds for the pseudometrics $\ell^c$ and $\overline{d}^c$ in the trajectory setting. The proof mostly relies on the technical elements prepared in the previous subsection. In addition, we need one more lemma.

%\paragraph*{Comparison to the fixpoint metric: }
%We are finally ready to compare the metric $\ell^c$ obtained through the logic $\cL_\sigma$ and $\cL_\tau$ to the metric $\overline{d}^c$ obtained through the functional $\cG_c$.

\begin{lem}
\label{lemma:mn-f}
For every $f \in \cL^c_\sigma$,  there exists $n \in \mathbb{N}$ such that $f \in \cH (d^c_n)$, i.e.,
\[\forall x,y\in E,\qquad  |f(x) - f(y)| \leq d^c_n(x,y).\]
For every $g \in \cL^c_\tau$,  there exists $n \in \mathbb{N}$ such that $g \in \cH(U_c(d^c_n))$, i.e.,
\[\forall\omega,\omega'\in\Omega,\qquad  |g(\omega) - g(\omega')| \leq U_c(d^c_n)(\omega, \omega').  \]
\end{lem}

\begin{proof}
The proof is done by induction on the structure of the logic $\cL_\sigma$ and $\cL_\tau$. We will only mention the cases that are different from the case of $\Lambda$ (see \cite{Chen25a}). 
% TODO: but the proof was not in the paper
\begin{itemize}
\item If $f = \int g$ with $g \in \cH(U_c(d^c_n))$, then
\begin{align*}
|f(x) - f(y)|
& = \left| \int g ~ \dee  \mathbb{P}^x - \int g~ \dee \mathbb{P}^y \right|\\
& \leq W(U_c(d^c_n)) (\mathbb{P}^x, \mathbb{P}^y)\\
& = \cG_c (d^c_n) (x, y) = d_{n+1}^c(x,y).
\end{align*}
\item If $g = \min \{ g_1, g_2\}$ with $g_i \in \cH(U_c(d_{n_i}^c))$ for $i=1,2$, then for every pair of trajectories $\omega, \omega'\in\Omega$, there is only one non-obvious case to study: $g_1(\omega) \leq g_2 (\omega)$ and $g_2(\omega') \leq g_1 (\omega')$. In this case,
\begin{align*}
|g(\omega) - g(\omega')|
& = |g_1(\omega) - g_2(\omega')|\\
& = \max \{g_1(\omega) - g_2(\omega'), g_2(\omega') - g_1(\omega) \}\\
&  \leq \max \{g_2(\omega) - g_2(\omega'), g_1(\omega') - g_1(\omega) \}\\
& \leq \max \{ U_c(d_{n_2}^c)(\omega, \omega'), U_c(d_{n_1}^c)(\omega, \omega')\}\\
& \leq U_c(d_{n}^c)(\omega, \omega')
\end{align*}
with $n := \max \{n_1, n_2\}$, which means that $g \in \cH(U_c(d_{n}^c))$.
\item If $g = f \circ ev_t$ with $f \in \cH(d_n^c)$, then for every pair of trajectories $\omega, \omega'\in\Omega$,
\begin{align*}
|g(\omega) - g(\omega')|
& =  \left|c^t f(\omega(t)) - c^t f(\omega'(t))\right|\\
& \leq c^t d_n^c (\omega(t), \omega'(t))\\
& \leq U_c(d_n^c)(\omega, \omega').
\end{align*}
\end{itemize}
\end{proof}

Finally, we obtain the equality of the pseudometrics $\ell^c$ and $\overline{d}^c$ (the analogue of Theorem~\ref{thm:F-logic}).

\begin{thm}\label{thm:G-logic}
The two pseudometrics $\ell^c$ and $ \overline{d}^c$ are equal.
\end{thm}

\begin{proof}
We know from Lemma \ref{lemma:comparison-fixpointG} and Theorem \ref{thm:mst-fixpoint} that $\ell^c \geq \overline{d}^c$. The other direction is a direct consequence of Lemma \ref{lemma:mn-f}.
\end{proof}

\color{black}
\begin{rem}
Each of the above behavioural pseudometrics
induces an equivalence relation: by Theorems \ref{thm:F-logic}
and \ref{thm:G-logic},
\[
\overline{\delta}^c(x,y)=0
\quad\Longleftrightarrow\quad
f(x)=f(y)\ \text{for every }f\in\grammar,
\]
and
\[
\overline{d}^c(x,y)=0
\quad\Longleftrightarrow\quad
f(x)=f(y)\ \text{for every }f\in\cL_\sigma^c.
\]
Thus, zero values of these pseudometrics
correspond to logical
indistinguishability.

However, we do not get immediate correspondence with the bisimulation notions introduced in \cite{Chen19a,Chen20,Chen23}. That is because those earlier works were based on indicator (and hence potentially discontinuous) observable functions,
whereas in this work we require $obs$ to be continuous. Linking our zero-distance relations to bisimulations
requires adapting the previous notions to continuous-$obs$ models and examining carefully the underlying
connections, which we leave for future work.
\end{rem}
\color{black}

\section{On the comparison between the two pseudometrics}
\label{sec:partial-conclusion}

In our previous work \cite{Chen19a,Chen20,Chen23}, we discussed both the importance and the challenge of involving trajectories in the study of behavioural equivalences for continuous-time processes. 

Let us first recall the example that led us to trajectories in \cite{Chen19a}: consider Brownian motion on $\mathbb{R}$ equipped with the observable $obs = \mathds{1}_{\{ 0\}}$ \footnote{This observable is not continuous and hence does not fit our current framework.}. The singleton $\{ 0\}$ is an $obs$-closed set\footnote{A Borel set $A\subseteq \mathbb{R}$ is \emph{$obs$-closed} when, for any $x,y\in\mathbb{R}$ with $obs(x)=obs(y)$, $x\in A$ if and only if $y\in A$.}, but for any $x \neq 0$ and any time $t$, $P_t(x, \{ 0\}) = 0$, which means that one cannot distinguish between the states $1$ and $1000$ when only looking at $P_t(\cdot, \{ 0\})$. However, invoking trajectories enables one to differentiate the non-zero states by tracking the hitting time of $\{0\}$ for the process starting from each state. In other words, the trajectories encode richer behavioural information about the process than the states. 

On the other hand, working with trajectories adds to the level of mathematical complexity in a considerable way, and certain notions do not translate easily from states to trajectories. For example, the natural analogue of an ``$obs$-closed'' set of states is what we called a \emph{time-$obs$-closed} set of trajectories, which is a measurable subset $B$ of the trajectory space such that, for every pair of trajectories $\omega, \omega'$ with $obs(\omega(t)) = obs(\omega'(t))$ for every time $t$, $\omega \in B$ if and only if $\omega' \in B$. However, the $\sigma$-algebra generated by the time-$obs$-closed sets cannot be simply described. 

For these reasons, we constructed in \cite{Chen25a} our first (set of) pseudometric $\overline{\delta}^c$ (and $\lambda^c$) only relying on the Markov kernels, while imposing a continuous $obs$ to prevent the ``obstacle'' encountered in the aforementioned Brownian motion example. In the current work, we manage to adapt this line of construction to trajectories in the case of diffusions, and acquire our second (set of) pseudometric $\overline{d}^c$ (and $\ell^c$). A natural question then arises: how these two pseudometrics compare with each other. 

To this end, we note that, by Corollary~\ref{cor:fixpoint-G-implies-F} and Theorem~\ref{thm:m-fixpointG}, for every $0<c<1$, the second pseudometric $\overline{d}^c$ is also a fixpoint for $\cF_c$ which is the functional that yields the first pseudometric $\overline{\delta}^c$. Hence, compiling the conclusions in Theorem~\ref{thm:F-fixpoint}, Theorem~\ref{thm:F-logic}, and Theorem~\ref{thm:G-logic}, we have the following order relation between the two (sets of) pseudometrics.
\begin{thm}\label{thm:pseudometric comparison}
    For every $0 < c < 1$,  $\lambda^c=\overline{\delta}^c \leq \overline{d}^c=\ell^c.$
\end{thm}

We will see with the concrete example studied in the next section that it is possible for the above inequality to hold strictly, which means that in general, the two pseudometrics are not exchangeable. However, as a potential direction for future study, it would be interesting to explore the conditions under which we can achieve $\overline{\delta}^c = \overline{d}^c$.  

\section{Example}
\label{sec:example}

Let us revisit one of the examples from \cite{Chen25a}.

We consider a process defined on $\{ 0, x, y, z, \partial\}$. Let us first give an intuition for what we are trying to model.  In the states $x$, $y$, $z$, the process is trying to learn a value. In the state $0$, the correct value has been learnt, but in the state $\partial$, an incorrect value has been learnt. From the three ``learning'' states $x, y$ and $z$, the process has very different learning strategies:
\begin{itemize}
\item from $x$,  the process exponentially decays to the correct value represented by the state $0$,
\item from $y$, the process is not even attempting to learn the correct value and thus remains in the same state, and
\item from $z$, the process slowly learns but it may either learn the correct value ($0$) or an incorrect one ($\partial$).
\end{itemize}
The word ``learning'' is here used only to give colour to the example.

Mathematically, this process is described by the time-indexed Markov kernels:
\begin{align*}
P_t (x, \{0\}) &= 1 - e^{- \lambda t} & P_t(z, \{ 0\}) &= \frac{1}{2}(1 - e^{- \lambda t}) & P_t(0, \{ 0\}) &= 1\\
P_t (x, \{x\}) &= e^{- \lambda t}  &  P_t(z, \{ \partial\}) &= \frac{1}{2}(1 - e^{- \lambda t}) &P_t(y, \{ y\})& = 1\\
&& P_t(z, \{ z\}) &=e^{- \lambda t} & P_t(\partial, \{ \partial\}) &= 1
\end{align*}
(where $\lambda > 0$)
and by the observable function $obs(x) = obs(y) = obs(z) = r \in (0,1)$, $obs(\partial) = 0$ and $obs(0) = 1$. We equip the state space with the discrete topology $\mathcal{O}$.

Note that in this case we can restrict our study to the following trajectories:
\begin{align*}
\Omega & = \Omega_x \cup \Omega_{z, 0} \cup \Omega_{z, \partial} \cup \{ \omega_\alpha ~|~ \alpha  \in \{ x, z, y, 0, \partial \} \}
\end{align*}
where $\omega_\alpha$ is the constant trajectory staying in $\alpha$ and
\begin{align*}
\Omega_x & := \{ \omega ~|~ \exists t>0 ~ \forall s < t ~ \omega (s) = x ~~ \text{and} ~~ \forall s \geq t ~ \omega(s) = 0 \},\\
\Omega_{z, 0} & := \{ \omega ~|~ \exists t>0 ~ \forall s < t ~ \omega (s) = z ~~ \text{and} ~~ \forall s \geq t ~ \omega(s) = 0 \},\\
\Omega_{z, \partial} & := \{ \omega ~|~ \exists t>0 ~ \forall s < t ~ \omega (s) = z ~~ \text{and} ~~ \forall s \geq t ~ \omega(s) = \partial \}.
\end{align*}
We can safely ignore $\omega_x$ and $\omega_z$ in subsequent computations as those trajectories have 0-probability.

%Those trajectories are not continuous in time. However, \textcolor{blue}{\sout{when restricted on the above specific sets of trajectories, no matter what pseudometric $m$ we have on the state space, $\cG_c(m)$ is continuous with respect to the topology generated by $m$.} (Linan: This is actually incorrect, as shown by the counterexample of the reviewer, because the restriction on subspaces of trajectories does not lead to a ``partition'' of the state space, and we still need to study $\overline d(x,y)$. I will add an argument in the appendix saying that the fixed point pseudometric $\overline d$ still exists in this setting.)}\\

\textcolor{black}{Those trajectories are not continuous in time. However, as we will explain in the  appendix (Section \ref{sec:appendix-example}), in this setting, the sequence $(d_n)_n$ still converges to a pseudometric $\overline d$ which is a fixed point for $\mathcal{G}_c$, and our comparison results on $\overline d$ and $\overline \delta$ still hold.}
%In this example the trajectories are not continuous, but are c\`adl\`ag and piecewise constant, so the trajectory space is Polish under the Skorohod metric. Since the state space is finite, and the dynamics are relatively simple, it is not hard to see that, if $m$ is a pseudometric on the state space that is continuous with respect to $\mathcal{O}$, so is $\mathcal{G}_c(m)$.
We pick the discount factor to be $c = e^{-\lambda} \in (0,1)$ and omit the superscript ``$c$'' in the discussions below.

Let us first recall the values of $\overline{\delta}$ we computed for this example in \cite{Chen25a}:
{ \small \[\begin{NiceArray}{|c|c|c|c|c|c|}
\hline
   & x & y & z & \partial & 0 \\ 
\hline
x & 0 & \frac{1 - r}{2} & \in \left[ \frac{1}{8}, \frac{1}{4} \right]  &  \begin{cases}
r ~~ \text{if } r \geq \frac{1}{2}\\
\frac{1}{2} ~~ \text{otherwise}
\end{cases} & 1 - r \\ \hline
y & \frac{1 - r}{2} & 0 & \frac{1}{4} & r & 1 - r \\ \hline
z &  \in \left[ \frac{1}{8}, \frac{1}{4} \right] & \frac{1}{4}& 0 & \begin{cases}
r ~~  \text{if } r \geq \frac{1}{4}\\
\frac{1}{4} ~~ \text{otherwise}
\end{cases} & \begin{cases}
\frac{1}{4} ~~ \text{if } r \geq \frac{3}{4}\\
1 - r ~~ \text{otherwise}
\end{cases} \\ \hline
\partial &  \begin{cases}
r ~~ \text{if } r \geq \frac{1}{2}\\
\frac{1}{2} ~~ \text{otherwise}
\end{cases} & r & \begin{cases}
r ~~  \text{if } r \geq \frac{1}{4}\\
\frac{1}{4} ~~ \text{otherwise}
\end{cases} & 0 & 1 \\ \hline
0 & 1 - r & 1 -r & \begin{cases}
\frac{1}{4} ~~ \text{if } r \geq \frac{3}{4}\\
1 - r ~~ \text{otherwise}
\end{cases} & 1 & 0 \\
\hline
\end{NiceArray} \] }

Switching to the pseudometric based on trajectories, we are able to determine the exact value for $\overline{d}$ for all pairs of states. We leave the detailed computations in the appendix (Section \ref{sec:appendix-example}).
{ \small \[\begin{NiceArray}{|c|c|c|c|c|c|}
\hline
   & x & y & z & \partial & 0 \\ 
\hline
x & 0 & 1 - r & 1  &  1 & 1 - r \\ \hline
y & 1 - r & 0 &  \max\{r, 1 - r\} & r & 1 - r \\ \hline
z & 1  &  \max\{r, 1 - r\} & 0 & 1 & 1 \\ \hline
\partial &  1 & r &1 & 0 & 1 \\ \hline
0 & 1 - r & 1 -r &1 & 1 & 0 \\
\hline
\end{NiceArray} \] }

This example shows that, compared to the first pseudometric $\overline{\delta}$, the second pseudometric $\overline{d}$ is very rigid and intolerant to differences. Since it is computed over the whole trajectories, $\overline{d}$ selects the worst case scenario and any deviation in behaviours will be exacerbated under $\overline{d}$. For instance, for the two states $z$ and $0$, $\overline{\delta}(z, 0)  =\max\{1-r,\frac{1}{4}\}$, while $\overline{d}(z, 0) = 1$. This discrepancy can be heuristically explained as follows: when comparing $z$ to $0$, our first pseudometric $\overline{\delta}$ is ``influenced'' by the fact that half of the weight on the state $z$ is going to shift over time to the state $0$; to the contrary, the second pseudometric $\overline{d}$ is ``dominated'' by the trajectories starting from $z$ going to the state $\partial$. This difference is further amplified by the inductive definition of $\overline{d}$.

Also note that this example is an illustration that $\overline{d} \geq \overline{\delta}$ (see Theorem \ref{thm:pseudometric comparison}) and the strict inequality can hold in certain occasions.

\section{Conclusion}
\label{sec:conclusion}

In the theoretical aspect, this work confirms that, similarly as in the case with discrete-time processes, proper behavioural pseudometrics for continuous-time processes can also be obtained via two equivalent iterative approaches -- through a functional and through a real-valued logic. Moreover, by following separately the perspective of the state space and that of the trajectory space, we acquire two different pseudometrics for continuous-time processes. We have established in Section \ref{sec:partial-conclusion} a general order relation between the two pseudometrics. However, more refined comparison results are still missing. In particular, it would be helpful to find out under what circumstances the two pseudometrics are equivalent.  

From the computational aspect, a drawback of our pseudometrics was already presented in the previous work \cite{Chen25a}, and that was, the exact computation of $\overline{\delta}$ was an involved process. Interlacing the transport theory with a supremum (over time) in the inductive definition of $\overline{\delta}$ made it virtually impossible to explicitly compute $\overline{\delta}$ for any hard example. Nonetheless, it was still possible to compare the behaviours of the process among given starting points. The computational difficulty posed by the transport theory and the Kantorovich metric also exists in the discrete-time case and there are interesting ways to handle it, for instance through the MICo distance \cite{Castro21}. 

\textcolor{black}{These difficulties concern explicit calculation rather
than computability in a formal sense. We do not establish
computability or complexity bounds here; such questions
require an effective presentation of the process and its
observable, together with tractable solutions to the concerned optimal transport
problems and their limit, and are beyond the
scope of this paper.}

The trajectory approach in the current work suffers from the same drawback that is further amplified by the fact that measures on the trajectory space are even less tractable than those on the state space. %the dynamics of the system is usually given by the time-indexed kernels instead of the probabilities on trajectories.
One potential avenue of solving this issue could be to consider alternatives to the supremums in time (in the definitions of both $\overline{d}$ and $\overline{\delta}$), such as certain integrals over time. Note that, in this case, the real-valued logics would also need to be adapted accordingly. The current logics both from \cite{Chen25a} and this work generalize the discrete-time logics very well. However, if pseudometrics are defined with integrals over time (instead of supremums), then it is unlikely that the corresponding real-valued logics relate that well to their discrete-time counterparts. 

\section*{Acknowledgment}

Linan Chen was supported by an NSERC discovery grant.

Florence Clerc was supported by the EPSRC-UKRI grant EP/Y000455/1, by a CREATE grant for the project INTER-MATH-AI, by NSERC and by IVADO through the DEEL Project CRDPJ 537462-18.

Prakash Panangaden was supported by an NSERC discovery grant.

\bibliographystyle{abbrv}
\bibliography{newmain}

@Book{Ash72,
  author =       "R. B. Ash",
  title =        "Real Analysis and Probability",
  publisher =    "Academic Press",
  year =         1972
}

@article{Bacci17,
  title={On-the-fly computation of bisimilarity distances},
  author={Bacci, Giorgio and Bacci, Giovanni and Larsen, Kim G and Mardare, Radu},
  journal={Logical Methods in Computer Science},
  volume={13},
  year={2017},
  publisher={Episciences. org}
}

@Book{Baier08,
  author =    {Christel Baier and Joost-Pieter Katoen},
  title =        {Principles of Model Checking},
  publisher =    {MIT Press},
  year =         2008}

@Book{Billingsley99,
  author =       {P. Billingsley},
  title =        {Convergence of Probability Measures},
  publisher =    {Wiley Interscience},
  year =         1999,
  edition =      {2nd}
}

@InProceedings{Blute97,
  author =       "R. Blute and J. Desharnais and A. Edalat and
  P. Panangaden",
  title =        "Bisimulation for Labelled {Markov} Processes",
  booktitle =    "Proceedings of the Twelfth IEEE Symposium On Logic In
  Computer Science, Warsaw, Poland.",
  year =         1997
}

@book{Bobrowski05,
  title={Functional analysis for probability and stochastic processes: an introduction},
  author={Adam Bobrowski},
  year={2005},
  publisher={Cambridge University Press}
}

@Book{Bourbaki89b,
  author =       {N. Bourbaki},
  title =        {General Topology Chapters 5-10},
  publisher =    {Springer-Verlag},
  year =         1989
}

@article{Castro21,
  title={{{M}{I}{C}o}: Improved representations via sampling-based state similarity for {M}arkov decision processes},
  author={Castro, Pablo Samuel and Kastner, Tyler and Panangaden, Prakash and Rowland, Mark},
  journal={Advances in Neural Information Processing Systems},
  volume={34},
  pages={30113--30126},
  year={2021}
}

@article{Chen19a,
  title = {Bisimulation for {{F}eller-{D}ynkin} Processes},
  journal = {Electronic Notes in Theoretical Computer Science},
  volume = {347},
  pages = {45 - 63},
  year = {2019},
  note = {Proceedings of the Thirty-Fifth Conference on the Mathematical Foundations of Programming Semantics},
  issn = {1571-0661},
  doi = {https://doi.org/10.1016/j.entcs.2019.09.004},
  author = {Linan Chen and Florence Clerc and Prakash Panangaden}
}

@article{Chen20,
title = "Towards a Classification of Behavioural Equivalences in Continuous-time {M}arkov Processes",
journal = "Electronic Notes in Theoretical Computer Science",
year = "2020",
note = "Proceedings of the Thirty-Sixth Conference on the Mathematical Foundations of Programming Semantics",
author = "Linan Chen and Florence Clerc and Prakash Panangaden",
}

@article{Chen23,
title={Behavioural equivalences for continuous-time {M}arkov processes}, 
DOI={10.1017/S0960129523000099}, 
journal={Mathematical Structures in Computer Science}, 
publisher={Cambridge University Press}, 
author={Chen, Linan and Clerc, Florence and Panangaden, Prakash}, 
year={2023}, 
pages={1–37}}

@InProceedings{Chen25a,
author="Chen, Linan
and Clerc, Florence
and Panangaden, Prakash",
editor="Abdulla, Parosh Aziz
and Kesner, Delia",
title="A Behavioural Pseudometric for Continuous-Time {M}arkov Processes",
booktitle="Foundations of Software Science and Computation Structures",
year="2025",
publisher="Springer Nature Switzerland",
address="Cham",
pages="24--44",
isbn="978-3-031-90897-2"
}

@Article{Desharnais02,
  author =       {J. Desharnais and A. Edalat and P. Panangaden},
  title =        {Bisimulation for Labeled {Markov} Processes},
  journal =      {Information and Computation},
  year =         2002,
  month =        "Dec",
  volume =       179,
  number =       2,
  pages =        {163-193}
}

@Article{Desharnais04,
  author =       {Jos\'ee Desharnais and Vineet Gupta and Radhakrishnan
                  Jagadeesan and Prakash Panangaden}, 
  title =        {A metric for labelled {Markov} processes},
  journal =      {Theoretical Computer Science},
  year =         2004,
  volume =       318,
  number =       3,
  pages =        {323-354},
  month =        {June}
}

@InProceedings{Desharnais98,
  author =       "J. Desharnais and A. Edalat and P. Panangaden",
  title =        "A Logical Characterization of Bisimulation for Labelled
  {Markov} Processes",
  pages =        "478-489",
  booktitle =    "proceedings of the 13th IEEE Symposium On Logic In
  Computer Science, Indianapolis",
  year =         1998,
  publisher =    "IEEE Press",
  month =        "June"
}

@InProceedings{Desharnais99b,
  author =       {J. Desharnais and V. Gupta and R. Jagadeesan and
  P. Panangaden},
  title =        {Metrics for Labeled {Markov} Systems},
  booktitle =    {Proceedings of CONCUR99},
  year =         1999,
  number =       1664,
  series =       {Lecture Notes in Computer Science},
  publisher =    {Springer-Verlag}
}

@Article{Ferns11,
  author =       {Norm Ferns and Prakash Panangaden and Doina Precup},
  title =        {Bisimulation Metrics for Continuous {M}arkov Decision Processes},
  journal =      {SIAM Journal of Computing},
  year =         2011,
  volume =    40,
  number =    6,
  pages =     {1662-1714}}

@InProceedings{Gupta04,
  author =       {Vineet Gupta and Radhakrishnan Jagadeesan and Prakash Panangaden}, 
  title =        {Approximate Reasoning for Real-Time Probabilistic Processes},
  booktitle =    {The Quantitative Evaluation of Systems, First International Conference QEST04},
  pages =        {304-313},
  year =         2004,
  publisher =    {IEEE Press}
}

@Article{Gupta06,
  author =       {Vineet Gupta and Radha Jagadeesan and Prakash Panangaden},
  title =        {Approximate reasoning for real-time probabilistic processes},
  journal =      {Logical Methods in Computer Science},
  year =         2006,
  volume =    2,
  number =    1,
  pages =     {paper 4}}

@book{Karatzas12,
  title={Brownian motion and stochastic calculus},
  author={Karatzas, Ioannis and Shreve, Steven},
  volume={113},
  year={2012},
  publisher={Springer Science and Business Media}
}

@Article{Larsen91,
  author =       {K. G. Larsen and A. Skou},
  title =        {Bisimulation through Probabilistic Testing},
  journal =      {Information and Computation},
  year =         1991,
  volume =       94,
  pages =        {1-28}
}

@BOOK{Milner80,
AUTHOR={R. Milner} ,
TITLE={A Calculus for Communicating Systems},
PUBLISHER={Springer-Verlag},
VOLUME={92},
SERIES={Lecture Notes in Computer Science},
YEAR={1980}
}

@article{Mustata08,
  author  = {Must{\u{a}}{\c{t}}a, Costic{\u{a}}},
  title   = {On a theorem of {Baire} about lower semicontinuous functions},
  journal = {Revue d'Analyse Num{\'e}rique et de Th{\'e}orie de l'Approximation},
  volume  = {37},
  number  = {1},
  pages   = {71--75},
  year    = {2008},
  doi     = {10.33993/jnaat371-877}
}

@InCollection{Park81,
  author =       {D. Park},
  title =        {Concurrency and automata on infinite sequences},
  booktitle =    {Proceedings of the 5th GI Conference on Theoretical
                  Computer Science},
  pages =        {167-183},
  publisher =    {Springer-Verlag},
  year =         1981,
  number =       104,
  series =       {Lecture Notes In Computer Science}
}

@Book{Rogers00a,
  author =    {L. Chris G. Rogers and David Williams},
  title =        {Diffusions, {M}arkov processes and martingales: {V}olume 1. {F}oundations},
  publisher =    {Cambridge university press},
  year =         2000,
  edition =   {2nd}}

@article{Sangiorgi09,
  title={On the origins of bisimulation and coinduction},
  author={Davide Sangiorgi},
  journal={ACM Transactions on Programming Languages and Systems (TOPLAS)},
  volume={31},
  number={4},
  pages={15},
  year={2009},
  publisher={ACM}
}

@article{vanBreugel05a,
author = 		{Franck van Breugel and James Worrell},
title = 		{A behavioural pseudometric for probabilistic transition systems},
journal = 		{Theoretical Computer Science},
year = 			{2005},
volume = 		{331},
number = 		{1},
pages = 		{115 - 142}
}

@book{Villani08,
  title={Optimal transport: old and new},
  author={C{\'e}dric Villani},
  year={2008},
  publisher={Springer-Verlag}
}

@Book{Whitt02,
  author =       {W. Whitt},
  title =        {An Introduction to Stochastic-Process Limits and their Applications to Queues},
  publisher =    {Springer-Verlag},
  year =         2002,
  series =       {Springer Series in Operations Research}
}

@Book{Williams91,
  author =       "David Williams",
  title =        "Probability with Martingales",
  publisher =    "CUP",
  year =         "1991",
  address =      "Cambridge",
  ISBN =         "0--521--40605--6",
  topic =        "probability theory",
}

\appendix

\appendix

\section{Rest of the computations for Section \ref{sec:example}}
\label{sec:appendix-example}

Note that in this case we can restrict our study to the following trajectories:
\begin{align*}
\Omega & = \Omega_x \cup \Omega_{z, 0} \cup \Omega_{z, \partial} \cup \{ \omega_\alpha ~|~ \alpha  \in \{ x, z, y, 0, \partial \} \}
\end{align*}
where $\omega_\alpha$ is the constant trajectory staying in $\alpha$ and
\begin{align*}
\Omega_x & := \{ \omega ~|~ \exists t>0 ~ \forall s < t ~ \omega (s) = x ~~ \text{and} ~~ \forall s \geq t ~ \omega(s) = 0 \},\\
\Omega_{z, 0} & := \{ \omega ~|~ \exists t>0 ~ \forall s < t ~ \omega (s) = z ~~ \text{and} ~~ \forall s \geq t ~ \omega(s) = 0 \},\\
\Omega_{z, \partial} & := \{ \omega ~|~ \exists t>0 ~ \forall s < t ~ \omega (s) = z ~~ \text{and} ~~ \forall s \geq t ~ \omega(s) = \partial \}.
\end{align*}
We can safely ignore $\omega_x$ and $\omega_z$ in subsequent computations as those trajectories have 0-probability.

We also define the random time $T: \Omega \to [0, + \infty]$ as follows. We define $T(\omega_\alpha) := + \infty$ for $\alpha  \in \{ x, z, y, 0, \partial \}$. 
If $\omega$ is in $\Omega_x \cup \Omega_{z, 0} \cup \Omega_{z, \partial}$, we define $T(\omega) := t$ where $t$ is the same as in the definition of those sets. Furthermore, we also define the random variable $X := \exp (- \lambda T)$. Note that $\mathbb{P}^x(X\leq q)=q$ for $q\in[0,1]$
and $\mathbb{P}^x(X\in\dee q)=\dee q$ on $(0,1)$
(and similarly for $\mathbb{P}^z$), since
$T$ has an exponential distribution with parameter $\lambda$.

\textcolor{black}{Under this setting, we can still define an increasing sequence of $1$-bounded pseudometrics $(d_n)_n$ with \[d_0(\alpha,\beta)=|obs(\alpha)-obs(\beta)|\text{ for }\alpha,\beta\in E,\quad\text{ and }d_{n+1}=\mathcal{G}_c(d_n)\text{ for }n\geq0,\]
and hence $\overline{d}=\lim_{n\to\infty}d_n=\sup_n\,d_n$ exists as a $1$-bounded pseudometric. Since the state space is finite, the convergence of $d_n$ to $\overline d$ is uniform, and $\mathcal{G}_c$ preserves this uniform convergence. To see the last statement, note that for every $\alpha,\beta\in E$ and $\gamma \in \Gamma(\mathbb{P}^\alpha, \mathbb{P}^\beta)$,
\begin{align*}
\int U_c(\overline d)\dee\gamma-\int U_c(d_{n})\dee\gamma&\leq\int|\sup_{t\ge0}c^t\,\overline d(\omega(t),\omega'(t))-\sup_{t\ge0}c^td_{n}(\omega(t),\omega'(t))|\dee\gamma\\
&\leq\int\sup_{t\geq0}c^t|\overline d(\omega(t),\omega'(t))-d_{n}(\omega(t),\omega'(t))|\dee\gamma\\
&\leq\sup_{\alpha',\beta'\in E}|\overline d(\alpha',\beta')-d_n(\alpha',\beta')|:=\|\overline d-d_n\|_\infty;
\end{align*}
taking the infimum over all $\gamma\in\Gamma(\mathbb{P}^\alpha, \mathbb{P}^\beta)$, we arrive at \[
|\mathcal{G}_c(\overline d)(\alpha,\beta)-\mathcal{G}_c(d_{n})(\alpha,\beta)|\leq \|\overline d-d_n\|_\infty,
\]
which guarantees that $\mathcal{G}_c(d_n)(\alpha,\beta)\to\mathcal{G}_c(\overline d)(\alpha,\beta)$. Finally, 
\[
|\mathcal{G}_c(\overline d)(\alpha,\beta)-\overline d(\alpha,\beta)|\leq |\mathcal{G}_c(\overline d)(\alpha,\beta)-\mathcal{G}_c(d_{n})(\alpha,\beta)|+|d_{n+1}(\alpha,\beta)-\overline d(\alpha,\beta)|\to0.
\]
We conclude that $\mathcal{G}_c(\overline d)=\overline d$, and hence our previous study of the fix point pseudometric $\overline d$ still applies in this example.}
%Note that those trajectories are not continuous in time. However, when restricted on the above specific sets of trajectories, no matter what pseudometric $m$ we have on the state space, $\cG_c(m)$ is continuous with respect to the topology generated by $m$. We can thus still apply our reasoning.
%as we will see we can still define the increasing sequence of distances $d_n$ and thus their supremum $\overline{d}$.  We will slightly abuse notations: instead of assuming that $d_n$ is constructed and then construct $d_{n+1}$ for every pair of points, we will look at pair of points $\alpha, \beta$ and assuming that $d_n(\alpha, \beta)$ is defined, construct $d_{n+1}(\alpha, \beta)$.

\begin{itemize}
\item First of all, note that for $\alpha, \beta \in \{ 0, y, \partial\}$,
\begin{align*}
\cG (d_0)(\alpha, \beta)
& = U_c(d_0)(\omega_\alpha, \omega_\beta)\\
& = d_0 (\alpha, \beta)
\end{align*}
and hence $\overline{d} (\alpha, \beta) = d_0 (\alpha, \beta)$.

\item Let us go on by considering $x$ and $0$. 
\begin{align*}
d_1 (x, 0)
& =  \cG(d_0) (x, 0) = W(U_c(d_0))(\mathbb{P}^x, \mathbb{P}^0)\\
& = \int U_c(d_0)(\omega, \omega_0) ~ \dee \mathbb{P}^x (\omega)\\
& = \int \sup_t ~ e^{- \lambda t} d_0(\omega(t), 0) ~ \dee \mathbb{P}^x (\omega)
\end{align*}
Now note that for a specific trajectory $\omega$ in $\Omega_x$, if $t < T(\omega)$, then $ d_0(\omega(t), 0) = d_0(x, 0) = 1 - r$ and for $t \geq T(\omega)$, $\omega(t) = 0$ and therefore $d_0(\omega(t), 0) = 0$. This means that,

\[d_1 (x, 0)= \int (1 - r) ~ \dee \mathbb{P}^x (\omega) = 1 - r\]
We can thus conclude that $\overline{d}(x, 0) = 1 - r$.

\item Now let us look at $x$ and $\partial$. Assume that we were able to define $d_n(x, \partial)$, then:
\begin{align*}
d_{n+1} (x, \partial)
%& = \cG(d_n) (x, \partial) = W(U_c(d_n))(\mathbb{P}^x, \mathbb{P}^\partial)\\
& = \int U_c(d_n)(\omega, \omega_\partial) ~ \dee \mathbb{P}^x (\omega)\\
%& = \int \sup_t ~ e^{- \lambda t} d_n(\omega(t), \omega_\partial(t)) ~ \dee \mathbb{P}^x (\omega)\\
& = \int \sup_t ~ e^{- \lambda t} d_n(\omega(t), \partial) ~ \dee \mathbb{P}^x (\omega)\\
& = \int \max \{ d_n (x, \partial), e^{- \lambda T(\omega)} d_n (0, \partial) \} ~ \dee \mathbb{P}^x (\omega)\\
& = \int \max \{ d_n (x, \partial),  X(\omega) \} ~ \dee \mathbb{P}^x (\omega)\\
& = \int \left[ d_n (x, \partial) \mathds{1}_{ d_n (x, \partial)\geq X (\omega) } + X(\omega)  \mathds{1}_{ d_n (x, \partial) < X (\omega) }\right] ~ \dee \mathbb{P}^x (\omega)\\
& = d_n (x, \partial) \mathbb{P}^x (d_n (x, \partial)\geq X) + \int_{d_n (x, \partial)}^1 q ~ \dee q\\
& = d_n (x, \partial)^2 + \frac{1}{2} (1 - d_n (x, \partial)^2)\\
& =  \frac{1}{2} (1 + d_n (x, \partial)^2)
\end{align*}
As one can see, we can define the increasing sequence $d_n(x, \partial)$ and its limit is $1$, meaning that $\overline{d}(x, \partial) = 1$.

\item Similarly, we can compute $\overline{d}(x, y) $. Assume that $ d_n (x, y) \leq 1-r$.
\begin{align*}
d_{n+1}(x, y)
%& = \int U_c(d_n)(\omega, \omega_y) ~ d \mathbb{P}^x (\omega)\\
& = \int \sup_t ~ e^{- \lambda t} d_n(\omega(t), y) ~ \dee \mathbb{P}^x (\omega)\\
%& = \int \max \{ d_n (x, y), e^{- \lambda T(\omega)} d_n (0, y) \} ~ \dee \mathbb{P}^x (\omega)\\
& = \int \max \{ d_n (x,y),   (1 - r) X(\omega) \} ~ \dee \mathbb{P}^x (\omega)
\end{align*}
Since $ d_n (x, y) \leq 1-r$, we can conclude that $ d_{n+1} (x, y) \leq 1-r$. Let us now finish the computation of $d_{n+1} (x, y)$ and $\overline{d} (x, y)$:
\begin{align*}
d_{n+1}(x, y)
& = \int \left[ d_n (x, y) \mathds{1}_{ d_n (x, y)\geq  (1 - r) X (\omega) } +  (1 - r) X(\omega)  \mathds{1}_{ d_n (x, y) <  (1 - r) X (\omega) } \right]~ \dee \mathbb{P}^x (\omega)\\
& = d_n (x, y) \mathbb{P}^x
(d_n (x, y)\geq (1 - r) X)
+ (1 - r) \int_{\frac{d_n(x,y)}{1-r}}^1 q ~ \dee q\\
& = \frac{d_n (x, y)^2}{1 - r}  + \frac{1-r}{2} \left(1 - \left( \frac{d_n (x, y)}{1-r}  \right)^2\right)\\
& =  \frac{1-r}{2} \left(1 + \left( \frac{d_n (x, y)}{1-r}  \right)^2\right)
\end{align*}
Now note that $d_0 (x,y) = 0$ and hence for every $n$, $d_n(x,y) \leq 1-r$.
We can thus define the increasing sequence $d_n(x, y) \leq 1-r$ and its limit is $1-r$, meaning that $\overline{d}(x, y) = 1-r$.

\item When looking at trajectories starting from $z$, one has to be careful whether said trajectory is in $\Omega_{z, 0}$ or $\Omega_{z, \partial}$.  Computations are otherwise quite similar:
\begin{align*}
d_{n+1} (z, 0)
& =  \int \sup_t ~ e^{- \lambda t} d_n(\omega(t), 0) ~ \dee \mathbb{P}^z (\omega)\\
& =  \int \left[ \mathds{1}_{\Omega_{z, 0}} (\omega) d_n(z, 0) +  \mathds{1}_{\Omega_{z, \partial}} (\omega) \max \{ d_n (z, 0), e^{- \lambda T(\omega)} \} \right]~ \dee \mathbb{P}^z (\omega)\\
\end{align*}
as for $\omega \in \Omega_{z, 0} \cup \Omega_{z, \partial}$
\[ d_n (\omega(t), 0) = 
\begin{cases}
d_n (z, 0) ~~ \text{if} ~~ t < T(\omega)\\
d_n (\partial, 0) = 1 ~~ \text{if}~~ t \geq T(\omega) ~\text{and} ~ \omega \in \Omega_{z, \partial}\\
d_n (0, 0) = 0 ~~ \text{if}~~ t \geq T(\omega) ~\text{and} ~ \omega \in \Omega_{z, 0}
\end{cases} \]
Then,
\begin{align*}
d_{n+1} (z, 0)
& =  d_n(z, 0) \mathbb{P}^z(\Omega_{z, 0}) +   \int_{\Omega_{z, \partial}}  \max \{ d_n (z, 0), X(\omega) \} ~ \dee \mathbb{P}^z (\omega)\\
%& = \frac{d_n(z, 0)}{2} + \int_{\Omega_{z, \partial}} \left( d_n(z, 0) \mathds{1}_{d_n(z, 0) \geq X} + X  \mathds{1}_{d_n(z, 0) < X} \right)  ~ \dee \mathbb{P}^z \\
& = \frac{d_n(z, 0)}{2} +  d_n(z, 0) \frac{1}{2}  \int \mathds{1}_{d_n(z, 0) \geq X}  ~ \dee \mathbb{P}^z + \frac{1}{2} \int X  \mathds{1}_{d_n(z, 0) < X}  ~ \dee \mathbb{P}^z \\
%& = \frac{d_n(z, 0)}{2} +   \frac{d_n(z, 0)^2}{2}  + \frac{1}{2} \int_{d_n(z, 0)}^1 q ~ \deeq \\
& = \frac{d_n(z, 0)}{2} +   \frac{d_n(z, 0)^2}{2}  + \frac{1}{4} \left( 1 - d_n(z, 0)^2 \right)\\
& =  \frac{1}{4} (d_n(z, 0) + 1)^2.
\end{align*}
As one can see, we can define the increasing sequence $d_n(z, 0)$ and its limit is $1$, meaning that $\overline{d}(z, 0) = 1$.

\item Moving on to $\overline{d}(z, \partial)$:
\begin{align*}
d_{n+1} (z, \partial)
& =  \int \sup_t ~ e^{- \lambda t} d_n(\omega(t), \partial) ~ \dee \mathbb{P}^z (\omega)\\
& =  \int \left[ \mathds{1}_{\Omega_{z, \partial}} (\omega) d_n(z, \partial) +  \mathds{1}_{\Omega_{z, 0}} (\omega) \max \{ d_n ( z, \partial), e^{- \lambda T(\omega)} \} \right]~ \dee \mathbb{P}^z (\omega)\\
\end{align*}
as for $\omega \in \Omega_{z, 0} \cup \Omega_{z, \partial}$
\[ d_n (\omega(t), \partial) = 
\begin{cases}
d_n (z, \partial) ~~ \text{if} ~~ t < T(\omega)\\
d_n (\partial, \partial) = 0 ~~ \text{if}~~ t \geq T(\omega) ~\text{and} ~ \omega \in \Omega_{z, \partial}\\
d_n (0, \partial) = 1 ~~ \text{if}~~ t \geq T(\omega) ~\text{and} ~ \omega \in \Omega_{z, 0}
\end{cases} \]
Then,
\begin{align*}
d_{n+1} (z, \partial)
& =  d_n(z, \partial) \mathbb{P}^z(\Omega_{z, \partial}) +   \int_{\Omega_{z, 0}}  \max \{ d_n (z,\partial), X(\omega) \} ~ \dee \mathbb{P}^z (\omega)\\
& = \frac{d_n(z, \partial)}{2} + \int_{\Omega_{z, 0}} \left( d_n(z, \partial) \mathds{1}_{d_n(z, \partial) \geq X} + X  \mathds{1}_{d_n(z, \partial) < X} \right)  ~ \dee \mathbb{P}^z \\
& = \frac{d_n(z, \partial)}{2} +  d_n(z, \partial) \frac{1}{2}  \int \mathds{1}_{d_n(z, \partial) \geq X} ~ \dee \mathbb{P}^z  + \frac{1}{2} \int X  \mathds{1}_{d_n(z, \partial) < X}  ~ \dee \mathbb{P}^z \\
& = \frac{d_n(z, \partial)}{2} +   \frac{d_n(z, \partial)^2}{2}  + \frac{1}{2} \int_{d_n(z, \partial)}^1 q ~ \dee q \\
& = \frac{d_n(z, \partial)}{2} +   \frac{d_n(z, \partial)^2}{2}  + \frac{1}{4} \left( 1 - d_n(z, \partial)^2 \right)\\
& =  \frac{1}{4} (d_n(z,\partial) + 1)^2.
\end{align*}
As one can see, we can define the increasing sequence $d_n(z, \partial)$ and its limit is $1$, meaning that $\overline{d}(z, \partial) = 1$.

\item For $\overline{d}(y, z)$:
\begin{align*}
d_{n+1} (y, z)
& =  \int \sup_t ~ e^{- \lambda t} d_n(y, \omega(t)) ~ \dee \mathbb{P}^z (\omega)\\
& =  \int \left[ \mathds{1}_{\Omega_{z,0}} \max \{ d_n (y,  z), (1 - r) e^{- \lambda T} \}  +  \mathds{1}_{\Omega_{z, \partial}} \max \{ d_n (y,  z),  r e^{- \lambda T} \}  \right]~ \dee \mathbb{P}^z \\
\end{align*}
as for $\omega \in \Omega_{z, 0} \cup \Omega_{z, \partial}$
\[ d_n (y, \omega(t)) = 
\begin{cases}
d_n (y,z) ~~ \text{if} ~~ t < T(\omega)\\
d_n (y, \partial) = r ~~ \text{if}~~ t \geq T(\omega) ~\text{and} ~ \omega \in \Omega_{z, \partial}\\
d_n (y, 0) =  1-r ~~ \text{if}~~ t \geq T(\omega) ~\text{and} ~ \omega \in \Omega_{z, 0}
\end{cases} \]
Then,
\begin{align*}
d_{n+1} (y,z)
& = \int_{\Omega_{z, 0}} \max \{ d_n (y,  z),  (1-r) X(\omega) \}  ~ \dee \mathbb{P}^z (\omega) 
+   \int_{\Omega_{z, \partial}} \max \{ d_n (y,  z),  r X(\omega) \}  ~ \dee \mathbb{P}^z (\omega)\\
& =  \int_{\Omega_{z, 0}} \left( d_n(y,z) \mathds{1}_{d_n(y,z) \geq (1-r) X} +(1 -r) X  \mathds{1}_{d_n(y,z) < (1-r) X} \right)  ~ \dee \mathbb{P}^z\\
& \qquad+ \int_{\Omega_{z, \partial}} \left( d_n(y,z) \mathds{1}_{d_n(y,z) \geq r X} + r X  \mathds{1}_{d_n(y,z) < r X} \right)  ~ \dee \mathbb{P}^z \\
\end{align*}
There are three cases to consider. First, if $r = 1/2$:
\begin{align*}
d_{n+1} (y,z)
& =  \int_{\Omega_{z, 0}} \left( d_n(y,z) \mathds{1}_{d_n(y,z) \geq \frac{1}{2} X} +\frac{1}{2} X  \mathds{1}_{d_n(y,z) < \frac{1}{2} X} \right)  ~ \dee \mathbb{P}^z\\
& \qquad+ \int_{\Omega_{z, \partial}} \left( d_n(y,z) \mathds{1}_{d_n(y,z) \geq \frac{1}{2} X} + \frac{1}{2} X  \mathds{1}_{d_n(y,z) < \frac{1}{2} X} \right)  ~ \dee \mathbb{P}^z \\
& = d_n(y,z)  \int_{\Omega_{z, 0} \cup \Omega_{z, \partial}}  \mathds{1}_{d_n(y,z) \geq \frac{1}{2} X} ~ \dee \mathbb{P}^z +\frac{1}{2}  \int_{\Omega_{z, 0} \cup \Omega_{z, \partial}}X  \mathds{1}_{d_n(y,z) < \frac{1}{2} X}  ~ \dee \mathbb{P}^z\\
\end{align*}
It is easy to see from this expression that if $d_n (y, z) \leq 1/2$, then $d_{n+1} (y,z) \leq 1/2$. Since $d_0(y, z) = 0$, we can thus conclude that for every $n$, $d_n (y, z) \leq 1/2$ and thus:
\begin{align*}
d_{n+1} (y,z)
& = d_n(y,z)  \int_{\Omega_{z, 0} \cup \Omega_{z, \partial}}  \mathds{1}_{d_n(y,z) \geq \frac{1}{2} X} ~ \dee \mathbb{P}^z +\frac{1}{2}  \int_{\Omega_{z, 0} \cup \Omega_{z, \partial}}X  \mathds{1}_{d_n(y,z) < \frac{1}{2} X}  ~ \dee \mathbb{P}^z\\
& = d_n(y,z) \mathbb{P}^z (2 d_n(y,z) \geq X) +\frac{1}{2}  \int_{2 d_n(y,z)}^1 q ~ \dee q\\
& = 2 d_n(y, z)^2+ \frac{1}{4} \left( 1 - (2 d_n(y, z))^2 \right)\\
& = \frac{1}{4} + d_n(y, z)^2.
\end{align*}
As one can see, we can define the increasing sequence $d_n(y, z)$. This sequence has a limit $l$ and its limit is $1/2$, meaning that $\overline{d}(y, z) = 1/2$ (when $r = 1/2$).

For the second case, assume that $r < 1/2$, i.e. $r < 1-r$. Remember that we had:
\begin{align*}
d_{n+1} (y,z)
& =  \int_{\Omega_{z, 0}} \left( d_n(y,z) \mathds{1}_{d_n(y,z) \geq (1-r) X} +(1 -r) X  \mathds{1}_{d_n(y,z) < (1-r) X} \right)  ~ \dee \mathbb{P}^z\\
& \qquad+ \int_{\Omega_{z, \partial}} \left( d_n(y,z) \mathds{1}_{d_n(y,z) \geq r X} + r X  \mathds{1}_{d_n(y,z) < r X} \right)  ~ \dee \mathbb{P}^z 
\end{align*}
It is important to note that if $d_n(y,z) \leq 1-r$, then $d_{n+1}(y, z) \leq 1-r$. In particular, since $d_0(y, z) = 0$, we know that for all $n$, $d_n(y,z) \leq 1-r$. 

Now there are only two cases to distinguish further: first, if $d_n(y,z) \leq r$, then
\begin{align*}
d_{n+1} (y,z)
& = \frac{d_n(y, z)^2}{2(1 - r)} + \frac{1-r}{2} \int_{\frac{d_n(y, z)}{1 -r}}^1 q ~ \dee q +  \frac{d_n(y, z)^2}{2r} + \frac{r}{2} \int_{\frac{d_n(y, z)}{r}}^1 q ~ \dee q\\
& = \frac{d_n(y, z)^2}{2(1 - r)} + \frac{1-r}{4} \left( 1 - \left[ \frac{d_n(y, z)}{1 -r}\right]^2 \right) +  \frac{d_n(y, z)^2}{2r} + \frac{r}{4}  \left( 1 - \left[ \frac{d_n(y, z)}{r}\right]^2 \right) \\
& = \frac{1}{4} + \frac{d_n(y, z)^2}{4} \left( \frac{1}{r} + \frac{1}{1-r} \right)\\
& = \frac{1}{4} + \frac{d_n(y, z)^2}{4r (1 - r)}.
\end{align*}
\textcolor{black}{We claim that $d_n(y,z)>r$ when $n$ is sufficiently large. Assume otherwise. Then the limit $\overline d$ of $(d_{n}(y,z))_n$ following the above recursion formula must satisfy $\overline d=\frac 14+\frac{\overline d^2}{4r(1-r)}$, which does not have any real root. Therefore, we know that this sequence would eventually reach the case $ r < d_n(y,z) \leq 1 - r$.} 

Let us study what happens once we reach that case: assume now that $ r < d_n(y,z) \leq 1 - r$. Then, note that given a trajectory $\omega \in \Omega_{z, \partial}$, one cannot have that $d_n(y, z) < r X (\omega)$. This means that
\begin{align*}
d_{n+1} (y,z)
& =  \int_{\Omega_{z, 0}} \left( d_n(y,z) \mathds{1}_{d_n(y,z) \geq (1-r) X} +(1 -r) X  \mathds{1}_{d_n(y,z) < (1-r) X} \right)  ~ \dee \mathbb{P}^z \\
& ~ + \int_{\Omega_{z, \partial}} d_n(y,z) ~ \dee \mathbb{P}^z  \\
& = \frac{d_n(y, z)^2}{2(1 - r)} + \frac{1-r}{2} \int_{\frac{d_n(y, z)}{1 -r}}^1 q ~ \dee q + \frac{d_n(y, z)}{2}\\
& = \frac{1-r}{4} + \frac{d_n(y, z)^2}{4 (1 - r)} + \frac{d_n(y, z)}{2}
\end{align*}
Note that it follows from this that if $d_n(y, z) \leq 1 - r$, then $d_{n+1} (y, z) \leq 1-r$ and since $d_0(y, z) = 0$, we get that for every $n$,  $d_n(y, z) \leq 1 - r$.
As usual, this defines an increasing sequence with limit $1 - r$ and from this we conclude that $\overline{d}(y, z) = 1 - r$.

The third case to consider is if $r > \frac{1}{2}$, but in that case we obtain $\overline{d}(y, z) = r$. 

To summarize all three cases: $\overline{d}(y, z) = \max\{r, 1 - r\}$.

\item Finally, we can  compute $\overline{d}(x, z)$ following a different route. Recall that $\overline{d}$ is a fixpoint of $\cG$, meaning that
\[ \overline{d}(x, z) = \inf_{\gamma \in \Gamma (\mathbb{P}^x, \mathbb{P}^z)} \int U_c(\overline{d}) ~ \dee \gamma. \]
For every $\gamma\in\Gamma(\mathbb{P}^x,\mathbb{P}^z)$,
\begin{align*}
\int U_c(\overline{d}) ~ \dee \gamma
 &= \int_{\Omega_x \times \Omega_{z, 0}} \sup_t c^t ~\overline{d} (\omega(t), \omega'(t)) ~ \dee \gamma (\omega, \omega')\\
 &\hspace{2cm}+ \int_{\Omega_x \times \Omega_{z, \partial}} \sup_t c^t ~ \overline{d} (\omega(t), \omega'(t)) ~ \dee \gamma (\omega, \omega').
\end{align*}
For the first integral, i.e.  $\omega \in \Omega_x$ and $ \omega' \in \Omega_{z, 0}$, it is clear that
%\[  \sup_t c^t ~ \overline{d} (\omega(t), \omega'(t)) =
%\begin{cases}
%\max \left\{ \overline{d}(x, z), c^{T(\omega)}  \overline{d}(0, z) \right\}  \qquad \text{if } T(\omega) \leq T(\omega')\\
%\max \left\{ \overline{d}(x, z), c^{T(\omega')}  \overline{d}(x, 0) \right\}  \qquad \text{if } T(\omega) > T(\omega')\\
%\end{cases} \]
%and thus
\begin{align}
\label{eq:zpartial}
 \int_{\Omega_x \times \Omega_{z, 0}} \sup_t c^t ~\overline{d} (\omega(t), \omega'(t)) ~ \dee \gamma (\omega, \omega') 
 & \geq \int_{\Omega_x \times \Omega_{z, 0}}  \overline{d}(\omega(0), \omega'(0)) ~ \dee \gamma\\
 &=\overline d(x,z)\mathbb{P}^z(\Omega_{z, 0})=\frac 12\overline d(x,z).
\end{align}
Similarly, we study the second integral: $\omega \in \Omega_x$ and $ \omega' \in \Omega_{z, \partial}$, we have that \textcolor{black}{
\[  \sup_t c^t ~ \overline{d} (\omega(t), \omega'(t)) =
\begin{cases}
\max \left\{ \overline{d}(x, z), c^{T(\omega)}  \overline{d}(0, z), c^{T(\omega')}  \overline{d}(0, \partial)  \right\}  \quad &\text{if } T(\omega) < T(\omega'),\\
\max \left\{ \overline{d}(x, z), c^{T(\omega')}  \overline{d}(0, \partial)  \right\}  \qquad &\text{if } T(\omega) =T(\omega'),\\
\max \left\{ \overline{d}(x, z), c^{T(\omega')}  \overline{d}(x, \partial) \right\}  \qquad &\text{if } T(\omega) > T(\omega').\\
\end{cases} \]}
Using the values we have already computed $\overline d(0,\partial)=\overline d(x,\partial)=\overline d(0,z)=1$, we obtain
%\begin{align*}
% & \int_{\Omega_x \times \Omega_{z, \partial}} \sup_t c^t~ \overline{d} (\omega(t), \omega'(t)) ~ \dee \gamma (\omega, \omega')\\
% & =  \int_{\Omega_x \times \Omega_{z, \partial}} \left[ \mathds{1}_{T(\omega) < T(\omega')} \max \left\{ \overline{d}(x, z), c^{T(\omega)}  \right\} +  \mathds{1}_{T(\omega) >T(\omega')} \max \left\{ \overline{d}(x, z), c^{T(\omega')} \right\} \right] ~ \dee \gamma (\omega, \omega ')\\
% & = \int_{\Omega_x \times \Omega_{z, \partial}} \max \left\{  \overline{d}(x, z),  \max \{ X(\omega), X(\omega') \} \right\} ~ \dee \gamma (\omega, \omega ')
% \end{align*}}
\begin{align*}
  \int_{\Omega_x \times \Omega_{z, \partial}} \sup_t c^t~ \overline{d} (\omega(t), \omega'(t)) ~ \dee \gamma (\omega, \omega')&\geq \int_{\Omega_x \times \Omega_{z, \partial}}\max \left\{  \overline{d}(x, z),  X(\omega') \right\} ~ \dee \gamma (\omega, \omega ') \\
  &=\int_{\Omega_{z,\partial}}\max\{\overline d(x,z),X(\omega')\}\dee \mathbb P^z(\omega')\\
  &=\frac 12 (\overline d(x,z))^2+\frac 14(1-(\overline d(x,z))^2).
\end{align*}
Combining with the first integral bound, we have \[
\int U_c(\overline{d}) ~ \dee \gamma\geq \frac 12\overline d(x,z)+\frac 12 (\overline d(x,z))^2+\frac 14(1-(\overline d(x,z))^2)=\frac 14(1+\overline d(x,z))^2.
\]
Taking the infimum over all $\gamma$, we arrive at \[
\overline d(x,z)\geq \frac 14(1+\overline d(x,z))^2,
\]
which is only possible if $\overline d(x,z)=1$.
\end{itemize}

\end{document}